\documentstyle[12pt,graphics]{article}
\newcommand{\A}{\alpha}
\newcommand{\B}{\beta}
\newcommand{\D}{\delta}
\newcommand{\DE}{\Delta}

\newcommand{\E}{\eta}

\newcommand{\T}{\theta}

\newcommand{\e}{\varepsilon}

\newcommand{\bea}{\begin{eqnarray}}
\newcommand{\be}{\begin{equation}}
\newcommand{\ee}{\end{equation}}
\newcommand{\ena}{\end{eqnarray}}
\newcommand{\beano}{\begin{eqnarray*}}
\newcommand{\enano}{\end{eqnarray*}}
\newcommand{\oo}{&&\hspace{-10mm}}
\newcommand{\hf}{\frac{1}{2}}

\newcommand{\bp}{{\B^2\over 8\pi}}
\newcommand{\bpp}{{\B^4\over 64\pi^2}}
\newcommand{\nn}{\nonumber \\ }
\newcommand{\del}{\partial}

\newcommand{\vs}[1]{\rule[- #1 mm]{0mm}{#1 mm}}

\newcommand{\cg}{{\mbox{$\cal{G}$}}}

\newcommand{\cw}{\mbox{$\cal{W}$}}

\font\fld=msbm10 at 12 pt

\newcommand{\fl}[1]{\mbox{\fld #1}}     
\newcommand{\NPB}[1]{{\it Nucl. Phys.} {\bf B#1}}
\newcommand{\PLB}[1]{{\it Phys. Lett.} {\bf B#1}}

\newcommand{\CMP}[1]{{\it Comm. Math. Phys.} {\bf #1}}

\begin{document}
\input epsf
\renewcommand{\thefootnote}{\fnsymbol{footnote}}
\newpage
\setcounter{page}{0}

\vs{10}
\begin{center}

{\Large {\sc \protect{Quantum Integrability of Coupled N=1 }\\
\protect{Super Sine/Sinh-Gordon Theories}\\ 
\protect{and the Lie Superalgebra $D(2,1;\A).$}}} \\[7pt]

\vs{2}

{\large \protect{Jonathan M.~Evans\footnote{e-mail:
J.M.Evans@damtp.cam.ac.uk}\footnote{Supported by a PPARC Advanced
Fellowship}}}, {\em DAMTP, University of Cambridge\footnote{Silver
Street, Cambridge CB3 9EW, UK.}} \\[2mm] {\large and} \\[2mm] {\large
\protect{Jens Ole Madsen\footnote{e-mail: madsen@fpaxp1.usc.es}}},
\protect{ {\em Universidade de Santiago de
Compostela}\footnote{Facultade de F{\'\i}sica, E-15706 Santiago de
Compostela, Spain}}

\end{center}
\vfill

\centerline{ {\bf Abstract}}
\noindent
We discuss certain integrable quantum field theories in (1+1)-dimensions
consisting of coupled sine/sinh-Gordon theories
with $N=1$ supersymmetry, positive kinetic energy, 
and bosonic potentials which are bounded from below.
We show that theories of this type can be constructed as Toda models
based on the exceptional affine Lie superalgebra $D(2,1;\A)^{(1)}$
(or on related algebras which can be obtained as various limits)
provided one adopts appropriate reality conditions for the fields.
In particular, there is a continuous family of such models in
which the couplings and mass ratios all depend on the parameter $\A$.
The structure of these models is analyzed in some detail at the
classical level, including the construction of conserved currents 
with spins up to 4.
We then show that these currents generalize to the quantum theory,
thus demonstrating quantum-integrability of the
models.

\vfill
\rightline{hep-th/9712227}
\rightline{US-FT/35-97}
\rightline{DAMTP/97-147}
\rightline{December 1997}

\setcounter{footnote}{0}
\renewcommand{\thefootnote}{\arabic{footnote}}
\newpage
\section{Introduction}

Of the many integrable quantum field theories in (1+1)-dimensions,
perhaps the most famous is the sine-Gordon model: a single scalar
field $\phi$ of mass $m$ with lagrangian
\be
\label{bsin}
L_{\rm sin} (\phi;m) = \del_+ \phi \, \del_- \phi + {m^2 \over 2} 
\cos 2 \phi  \, .
\ee
A close relative which is slightly less well-known is the sinh-Gordon
model, with lagrangian
\be
\label{bsinh}
L_{\rm sinh} (\phi;m) = \del_+ \phi \, \del_- \phi - {m^2 \over 2} 
\cosh 2 \phi \, .
\ee
The two are related at a formal level by 
\be
\label{twist} 
L_{\rm sin} (\phi ; m) = - L_{\rm sinh} ( i \phi ; m) 
\ee
but despite this they have quite different properties. 
The sinh-Gordon theory is actually significantly simpler, involving
a single species of particle created by the field $\phi$ whose 
quantum scattering is given by an elastic S-matrix. 
The sine-Gordon model, by contrast, has a potential with degenerate vacua and
hence solitons which interpolate between these. The `fundamental'
meson corresponding to $\phi$ now
emerges as a bound state which may even disappear from the
spectrum completely for certain values of the coupling
(we have suppressed the dimensionless coupling constants above).

The sine/sinh-Gordon models are 
merely the simplest members 
of a general class of
massive integrable field theories, known as Toda theories, 
which can be constructed from affine Lie algebras. The simple roots
can be used to define exponential interactions between scalar fields
living in the Cartan subalgebra so as to ensure the existence of a Lax pair 
which then guarantees classical integrability.
The reality properties of the fields can be
changed by a trick analogous to (\ref{twist}) whenever 
there is a reflection symmetry 
of the affine Dynkin diagram, 
with different choices of reality conditions essentially corresponding
to different real forms of a given complex algebra.
In the simplest cases above, the algebra is of complex 
type $A^{(1)}_1$, with the 
sine-Gordon and sinh-Gordon theories corresponding to the 
real forms $su(2)^{(1)}$ and $sl(2)^{(1)}$ respectively.

Supersymmetric versions of (\ref{bsin}) and (\ref{bsinh}) are readily
constructed. The $N=1$ sine-Gordon model can be defined in superspace by a
lagrangian 
\be \label{ssin} 
{\cal L}_{\rm sin} (\Phi;m) = 
{i \over 2} D_+ \Phi D_- \Phi + m \cos \Phi 
\ee
where $\Phi$ is a real superfield.\footnote{Our superspace conventions
will be discussed in detail in section 3; the particular numerical
coefficients appearing in the bosonic and superspace lagrangians are
consistent with the expansion of superfields in components that we
shall consider later.}
Similarly, the $N=1$ sinh-Gordon theory is
\be \label{ssinh} 
{\cal L}_{\rm sinh} (\Phi;m) = 
{i \over 2} D_+ \Phi D_- \Phi - m \cosh \Phi 
\ee
It is not surprising that these generalizations exist, since any 
theory of a single scalar field with a bounded potential
admits a supersymmetric extension.
But for other Toda models, with more than one field, the situation is
rather subtle. The most systematic approach is to generalize the
Lax-pair construction, replacing the underlying affine Lie algebra
with an affine Lie superalgebra (see \cite{EM} for a review). 
Even then the theory is 
supersymmetric only if the system of simple roots used is totally
fermionic. The bosonic sector of such a model does not consist 
simply of a single 
bosonic Toda theory, but rather of a number of different
bosonic Toda theories which are related to the choice of superalgebra 
simple roots.
In other words, it is only these specific combinations of bosonic Toda
models which admit supersymmetric extensions\footnote{
It is, however, possible to construct more general theories with non-linearly 
realized supersymmetry, see \cite{To}}.

Just as in the bosonic situation, the superspace lagrangians (\ref{ssin}) and
(\ref{ssinh}) are related by
\be
\label{stwist} 
{\cal L}_{\rm sin} (\Phi ; m) = - {\cal L}_{\rm sinh} ( i \Phi ; m) 
\ee
but the idea of changing reality conditions in this way 
actually takes on a new significance for 
superalgebras. 
For bosonic Toda models, a maximally
non-compact real algebra defines a 
theory with positive-definite kinetic energy, 
while other real forms produce theories with indefinite kinetic
energy whose physical status is therefore less clear. The single
exception to this is the sine/sinh-Gordon correspondence above, where
the new lagrangian obtained from the substitution $\phi\rightarrow i\phi$ 
happens to be negative-definite, rather than
indefinite, so that it is still physically sensible. 
For superalgebra Toda models, however, there are a number of more
complicated ways to generalize the behaviour (\ref{stwist}) and still
arrive at physically sensible theories. In a sense, the bosonic
situation is reversed, because the 
maximally non-compact form of a Lie superalgebra 
typically gives rise to a Toda model with indefinite kinetic energy,
which can then be rendered 
positive-(or negative-)definite in a limited number of cases 
by an appropriate change of reality condition.

The first systematic analysis of affine superalgebra theories along
these lines was recently carried out in \cite{EM}, to which we refer the 
reader for additional background, including more detailed references to
earlier work.
A number of cases emerged as being of special interest,
and it is our aim in this paper to study the most novel and important 
of these, corresponding to the exceptional Lie superalgebra
$D(2,1;\A)^{(1)}$ (the others can actually be recovered in various limits).
This gives a
one-parameter family of massive, supersymmetric theories with positive
kinetic energy, whose bosonic sectors consist of two sine-Gordon
models and one sinh-Gordon model,  
with relations between the masses and
couplings which are fixed by the value of the
parameter $\A$. 
It is certainly not obvious that these particular combinations of 
sine/sinh-Gordon theories can be made to interact in a supersymmetric
fashion while maintaining 
classical integrability, but all this is
guaranteed by the superalgebra construction. 
We shall show at the end of the paper that integrability
of this family of models holds quantum-mechanically too.

We give a brief introduction to
the Lie superalgebra
$D(2,1;\A)$ and its affine extension in section \ref{d21}, which is 
sufficient to allow us to write down the classical theory. 
Thereafter we shall
study the model, including its integrability properties, 
directly from its lagrangian.
In section \ref{toda} we analyze the
classical theory in some detail, describing
how to `twist' it along the lines of (\ref{twist}) and (\ref{stwist}) 
to ensure that it has positive kinetic energy, 
and showing that the resulting bosonic sub-sector of the model 
is a sum of two sine-Gordon theories and one sinh-Gordon theory.
We find the classical masses and
three-point functions of the various particles, and describe the 
solitonic solutions to the equations of motion.
In section \ref{other} we discuss other models which can be obtained 
from the 
$D(2,1;\A)^{(1)}$ model in various limits. 
In section \ref{ccons} we construct
explicitly all classically conserved currents 
of (super)spins up to 7/2 in the $D(2,1;\A)^{(1)}$ family, and   
in section \ref{qcons} we show that these currents 
survive quantization, with suitable modifications which we calculate, thus
demonstrating quantum integrability of the theory.
Section \ref{con} summarizes our results and discusses some 
projects for future study. 

\section{The Exceptional Lie Superalgebra $D(2,1;\A)$} 
\label{d21}

We recall that a Lie superalgebra $\cg$ is a $\fl{Z}_2$-graded algebra
$\cg=\cg_0\oplus\cg_1$ equipped with a graded commutator satisfying the 
graded Jacobi-identity. The even subspace $\cg_0$ is a Lie algebra 
(non-semisimple in general), while the odd subspace $\cg_1$ forms a 
representation of $\cg_0$. 
The basic (or contragredient) Lie superalgebras can be characterized by a
Cartan matrix $c_{ij}$ corresponding to a set of simple roots $\A_i$. 
A root $\A_i$ is called bosonic (fermionic) if the corresponding generator 
$E_{\A_i}$ belongs to the even (odd) subspace. 
The basic Lie superalgebras are equipped with an invariant bilinear 
form which induces an inner-product
on the root-space which is generally of indefinite signature. 
Bosonic roots can thus have positive 
or negative length squared, while fermionic roots can in addition have 
zero length squared. 
In the Dynkin diagram for a Lie superalgebra $\cg$ the bosonic roots are 
represented by ``white'' nodes {\Large $\circ$}, 
null fermionic roots (fermionic roots with zero length) 
are represented by ``grey'' nodes 
$\hbox{\epsfxsize=7pt \epsfbox{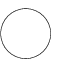}}$, and fermionic roots with non-zero 
length are represented by ``black'' nodes {\Large $\bullet$}. 
Note that, in contrast to the case of Lie algebras, for Lie superalgebras 
the Cartan matrix is not uniquely 
defined; there are in general several possible inequivalent choices of simple 
roots. For more information on Lie superalgebras see e.g. \cite{EM,Dict}. 

The exceptional basic Lie superalgebra $D(2,1;\A)$ is a deformation of
the Lie superalgebra $D(2,1) = osp(4 , 2)$, 
with a continuous deformation parameter
$\A$; $D(2,1)$ corresponds to taking $\A=1$ (or $\A=-2$, or
$\A=-\hf$, as we will see shortly). In general $\A$ takes values in 
$\fl{C}\setminus \{-1,0\}$, but we will consider only 
$\A\in\fl{R}\setminus\{-1,0\}$, since only in this case do we get a 
{\em real} inner-product on the root-space. 
The even (bosonic) subalgebra of $D(2,1;\A)$ is 
$\cg_0 = A_1\oplus A_1\oplus A_1$, while the 
odd subspace transforms as the (2,2,2) representation of $\cg_0$. 

The simple root system of 
$D(2,1;\A)$ which will be relevant for us consists of fermionic roots
$\A_i$ $(i=1,2,3)$, spanning a three dimensional vector-space $V$.  The
simple roots are all null, $\A_i^2 = 0$, but they have
non-vanishing inner-products with each other:
\be 
\A_1\cdot\A_2 = \A; 
\qquad \A_1\cdot\A_3 = 1; 
\qquad \A_2\cdot\A_3 = -1-\A
\ee
A convenient description can be
given using orthogonal basis vectors $\e_i$ $(i=1,2,3)$ 
for $V$, normalized so that
\be 
\e_1^2 = -{1+\A \over 2};\qquad 
\e_2^2 =  \hf;\qquad 
\e_3^2 = {\A\over 2} \, .
\ee   
In terms of these we have 
\be 
\A_1 = -\e_1+\e_2+\e_3; \quad 
\A_2 =  \e_1-\e_2+\e_3; \quad
\A_3 = \e_1+\e_2-\e_3 \, . 
\ee 
The full root system of $D(2,1;\A)$ contains fermionic elements 
$\{ \pm \e_1 \pm \e_2 \pm \e_3\}$ (all combinations of signs allowed) 
and bosonic elements $\{ \pm 2\e_i \}$.
The Cartan matrix and Dynkin diagram are given by:\\[7pt]
\centerline{
\hbox{ 
\raise 20pt \hbox{$c_{ij} = \left (
\begin{array}{ccc} 0 & \A & 1 \\ \A & 0 & -1-\A \\
1 & -1-\A &     0 
\end{array} \right )$} \hspace{15mm}
\vbox{\hbox{\epsfxsize=60pt \epsfbox{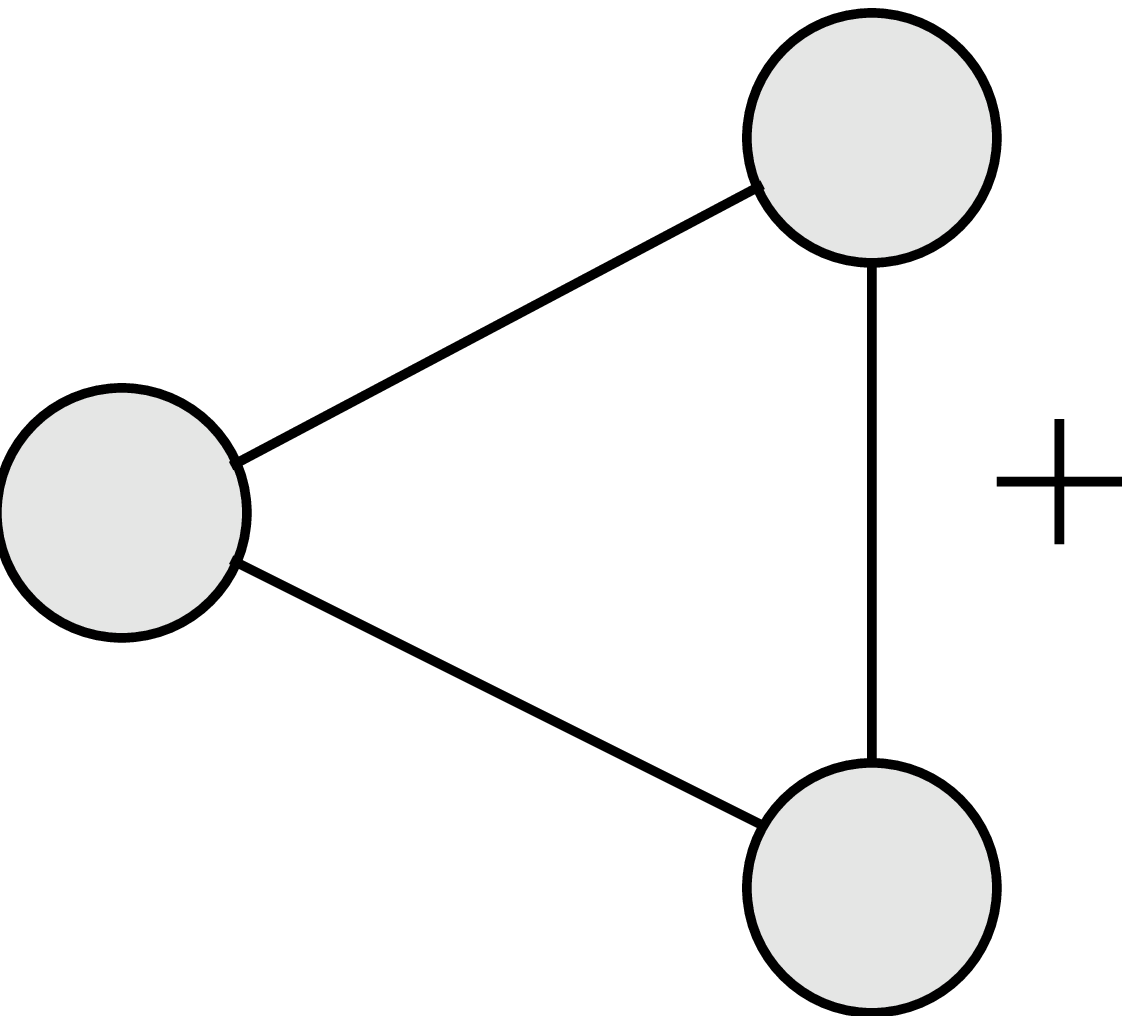} 
\hskip -52pt 
\raise 1pt \vbox{\hbox{\hskip 17pt $\A$} \hbox{\bigskip} 
\hbox{--1--\hskip 1pt $\A$}}}}
}}\\[7pt]
Note that there is no way to normalize the Cartan matrix such that the entries 
are integers. In order to draw the Dynkin diagram we therefore cannot
follow the usual procedure, 
to draw $n_{ij}$ ``bonds'' between nodes $i$ and $j$, 
where $n_{ij} = {\rm max}(|c_{ij}|,|c_{ji}|)$. Instead we draw a bond between 
nodes $i$ and $j$, and we write the actual value 
of $c_{ij}=c_{ji}$ on the bond. 
The ``+'' in the diagram stands for a value of +1.  

Replacing $\A$ by $-1-\A$ in the Cartan matrix
is equivalent to a reordering of the
simple roots; and replacing $\A$ by ${1/\A}$ is again equivalent to
such a reordering, after multiplying the matrix by an overall factor $\A$. 
We conclude that there is an isomorphism between 
algebras with
parameters $\A,\,-1-\A$ and ${1\over\A}$; in fact the transformations
$\A \rightarrow -1-\A$ and $\A \rightarrow {1\over\A}$ generate a
group of order 6 (permutations of 3 objects) 
and the following parameters all correspond to 
isomorphic algebras:
\be
\label{iso}
\A \sim -1-\A \sim {1\over \A} \sim -{1\over 1+\A} \sim 
-{\A\over 1+\A} \sim -{1+\A\over\A}
\ee
Note that $D(2,1;-2)$ and $D(2,1;-\hf)$ are both isomorphic to 
$D(2,1;1)$ and thus to $D(2,1)$. 
The equivalences (\ref{iso}) connect the three intervals of real numbers
$\,(\!-\!\infty,-1)$, $(\!-\!1,0)$ and $(0,\infty)$.
This allows us to restrict attention to $\A$ positive if we so wish,
since an algebra $D(2,1;\A)$ with a negative parameter
$\A$ is always isomorphic to an algebra with a positive value of $\A$. 

To pass from $D(2,1;\A)$ to its affine extension $D(2,1;\A)^{(1)}$,  
we must add an additional generator of non-zero grade, rendering the 
algebra infinite-dimensional. For our purposes it is sufficient to 
consider the projection of the new root system onto the horizontal or
zero-grade subspace spanned by the original simple roots of $D(2,1;\A)$. 
In this subspace, the projection of the affine root 
is simply the lowest root of 
$D(2,1;\A)$, namely 
\be \A_0 \quad = \quad  -(\A_1+\A_2+\A_3) \quad = \quad 
- (\e_1 + \e_2 + \e_3 )
\ee
The Cartan matrix and Dynkin diagram of the affine Lie algebra are: 
\be\label{DD}
\hbox{\raise 20pt 
\hbox{$
\left ( \begin{array}{cccc} 
0 & -1-\A & 1  & \A \\ -1-\A & 0 & \A & 1 \\ 1 & \A & 0 & -1-\A \\
\A & 1 & -1-\A & 0
\end{array} \right )$} \hspace{15mm}
\hbox{\kern 30pt \vbox{\hbox{\medskip}\hbox{\epsfxsize=55pt 
\epsfbox{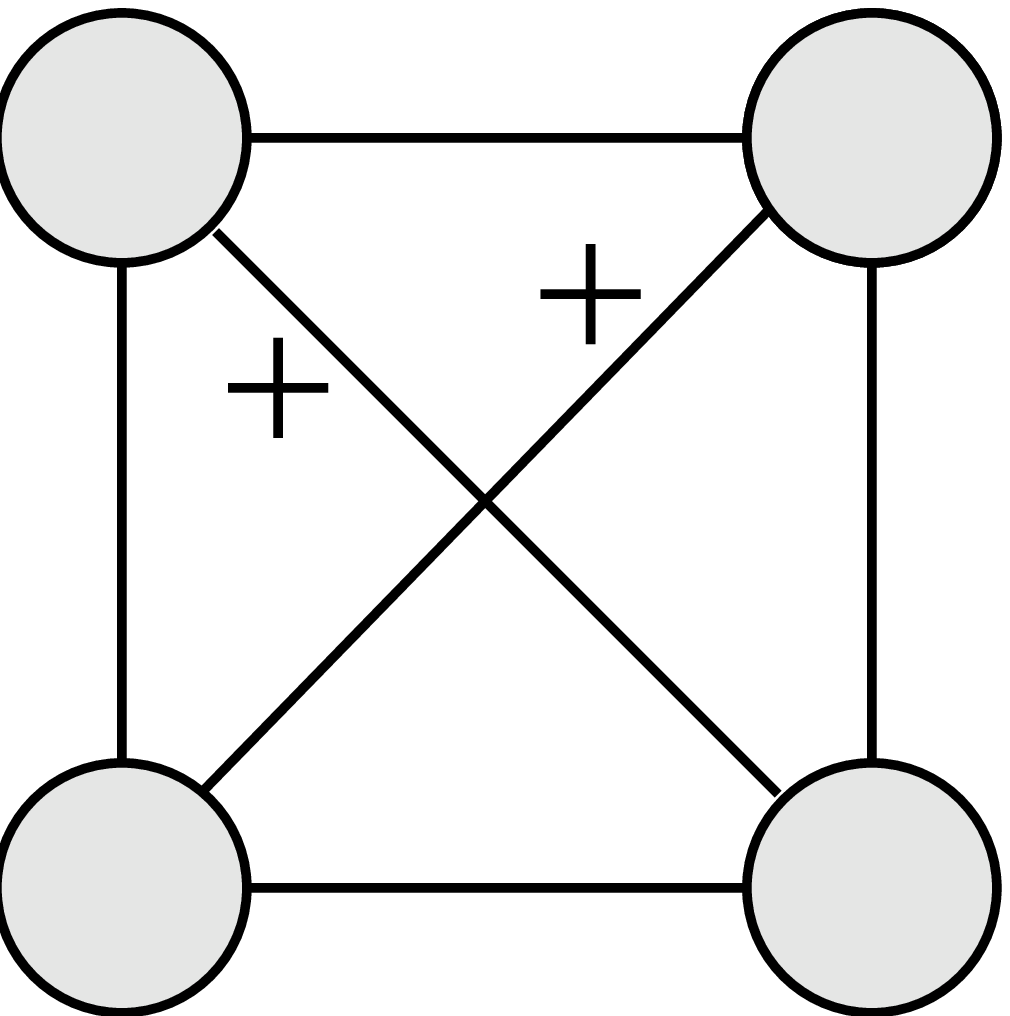} 
\raise 25pt \hbox{\hskip -78pt --1--\hskip 1pt $\A$} 
\raise 52pt \hbox{\hskip 18pt $\A$}
\raise -2pt \hbox{\hskip -12pt $\A$}
\raise 25pt \hbox{\hskip 14pt --1--\hskip 1pt$\A$ \hskip 0pt}}}}
}
\ee

\section{The Toda Models Defined by $D(2,1;\A)^{(1)}$} 
\label{toda}

\subsection{Standard Reality Conditions}

To write down an integrable model corresponding to a
finite-dimensional Lie
superalgebra $\cg$ or its affine extension $\cg^{(1)}$, 
we follow the general procedure
summarized in \cite{EM}.  We work in two-dimensional super-Minkowski
space with bosonic light-cone coordinates $x^\pm$ and fermionic
light-cone coordinates $\theta^\pm$ which are real (or Majorana)
spinors; covariant super-derivatives are defined by $$ D_{\pm}= {\del
\over \del
\theta^{\pm}} - i \theta^{\pm} \del_{\pm} \, , \quad 
D^2_\pm = - i \del_\pm \ , \quad \{ D_+ , D_-\} = 0 \ .  $$
We introduce a superfield $\Phi$ with values in the real vector space
$V$ of dimension rank$(\cg)$ 
spanned by the simple roots, which we assume are all odd. 
An integrable and manifestly supersymmetric theory
is then defined by a superspace lagrangian density 
\be
{\cal L} = \frac{i}{2} D_+ \Phi \cdot D_- \Phi + \sum_j\mu\exp\A_j\cdot\Phi
\ee
where $\mu$ is a parameter with the dimensions of mass. 
It is also
customary to introduce a dimensionless coupling constant $\B$ in the
theory as an overall factor ${1/\B^2}$ multiplying the lagrangian,
but we will refrain from doing this until we discuss quantization 
in section 6. 
The superfield equations of motion following from this lagrangian are 
\be
iD_+ D_- \Phi = \sum_{j} \mu \A_j\exp{\A_j\cdot\Phi} \, .
\ee

For a finite-dimensional algebra $\cg$, the 
sums above run over
$j = 1, \ldots , {\rm rank}(\cg)$. This type of model is conformally
invariant and the mass parameter $\mu$ 
can actually be removed by a shift in the
the superfield $\Phi$. The Toda theory based on the affine algebra 
$\cg^{(1)}$ has an additional term in the potential labelled by $j =
0$ corresponding to the affine root, though the fields live in the same
space $V$. This additional term gives a minimum to the potential,
breaking conformal invariance.
In either case 
the equations of motion can be written as a zero-curvature condition 
for a Lax connection in superspace and this is what guarantees the classical 
integrability of the theory. 
Although we are interested primarily in massive theories, some
knowledge of the conformal models will be important later.

The case of interest to us is $\cg = D(2,1;\A)$ and to write down the
models explicitly we define a convenient 
set of fields
\be \Phi_i = \e_i\cdot\Phi \ee 
where $\e_i$ are the orthogonal basis vectors introduced in the last
section.
It is also useful to introduce the following notation for terms in the
potential:
$$U_j (\Phi) = \exp (\alpha_j \cdot \Phi)$$
or in detail:
\bea
U_0 (\Phi) & = & \exp(-\Phi_1-\Phi_2-\Phi_3) \nn
U_1 (\Phi) & = & \exp(-\Phi_1+\Phi_2+\Phi_3) \nn
U_2 (\Phi) & = & \exp(+\Phi_1-\Phi_2+\Phi_3) \nn
U_3 (\Phi) & = & \exp(+\Phi_1+\Phi_2-\Phi_3) 
\ena 
We can then write the superspace lagrangian for either the conformal
or massive models as
\be
\label{susy}
{\cal L} = {-i\over 1+\A} D_+\Phi_1 D_-\Phi_1 + i D_+\Phi_2
D_-\Phi_2 + {i\over\A}D_+\Phi_3 D_-\Phi_3 + \mu U(\Phi) \ .
\ee
For the conformal theory based on $D(2,1;\A)$
the potential term is: 
$$
U(\Phi) = U_1 (\Phi) + U_2 (\Phi) + U_3 (\Phi) \ .
$$
For the massive theory based on the affine algebra  
$D(2,1;\A)^{(1)}$ the potential is
\bea 
U(\Phi) & = & U_0 (\Phi) + U_1 (\Phi) + U_2 (\Phi) + U_3 (\Phi) \nn
&=& 4\left ( \cosh\Phi_1 \cosh\Phi_2 \cosh\Phi_3 -
\sinh\Phi_1\sinh\Phi_2\sinh\Phi_3 \right ). 
\ena 
The equations of motion for the massive model are then:
\bea
\label{eom}
2iD_+ D_- \Phi_1 & = & -\mu(1+\A)\left ( - U_0 - U_1 + U_2 + U_3 \right ) \nn 
2iD_+ D_- \Phi_2 & = & \mu\left ( - U_0 + U_1 - U_2 + U_3 \right ) \nn 
2iD_+ D_- \Phi_3 & = & \mu\A\left ( - U_0 + U_1 + U_2 - U_3 \right ) 
\ena
Note that in either case the lagrangian depends on the parameter 
$\A$ through its appearance in 
the inner-product on the target-space $V$; in fact $\A$ can be
regarded as an additional 
dimensionless coupling constant. 

To understand better the structure of the massive theory, we 
can check from the equations of motion 
that any two of the three fields can consistently be set to zero. 
On doing so, the resulting single-field Lagrangians are:
\bea
{\cal L}_{\Phi_2=\Phi_3=0} &=& -{1\over 1+\A} \left ( 
i D_+\Phi_1 D_- \Phi_1 - 4 \mu(1+\A) \cosh\Phi_1 \right ) \label{ssg1}
\\
{\cal L}_{\Phi_1=\Phi_3=0} &=& \left (
i D_+\Phi_2 D_- \Phi_2 + 4 \mu \cosh\Phi_2 \right ) \label{ssg2} \\
{\cal L}_{\Phi_1=\Phi_2=0} &=& {1\over \A} \left ( 
iD_+\Phi_3 D_- \Phi_3 + 4 \mu\A \cosh\Phi_3 \right ) \label{ssg3}
\ena 
Comparing with (\ref{ssinh}) we see that 
these are just three $N=1$ sinh-Gordon theories with masses $2\mu(1+\alpha)$, 
$-2\mu$ and $-2\mu\A$ respectively, and with a relative minus sign between the 
first and the last two sinh-Gordon lagrangians. 

To gain more insight, we pass to the component formulation of the
theory by expanding the superfields:
$$
\Phi=\phi + i \theta^+ \psi_{+}  + i \theta^- \psi_{-} + 
i \theta^+ \theta^- \sigma \ .  
$$
After eliminating the auxiliary fields in the usual way
we obtain a component-field lagrangian of the form
$$ 
L = L_{\rm bos} (\phi) + L_{\rm ferm} (\psi) + L_{\rm int} (\phi, \psi) , 
$$ 
where the various terms appearing have the following structure:
The bosonic sector takes the form
\bea
\label{sinh}
L_{\rm bos} & = & -{1\over 1+\A} \left (
\del_+ \phi_1 \del_- \phi_1 - 2\mu^2(1+\A)^2 \cosh 2\phi_1 \right ) \nn
&& \hfill 
+ \quad \left ( \del_+ \phi_2 \del_- \phi_2 - 2\mu^2 \cosh 2\phi_2 \right) \nn
&& \hfill 
+ \quad {1\over\A}\left (
\del_+ \phi_3 \del_- \phi_3 - 2\mu^2\A^2 \cosh 2\phi_3 \right ) \nn
& = & -{1\over 1+\A} L_{\rm sinh}(\phi_1;m(1+\A))+L_{\rm sinh}(\phi_2;m)+
{1\over\A}L_{\rm sinh}(\phi_3;m\A)
\ena
where $L_{\rm sinh}$ is defined in (\ref{bsinh}),
and we have set $m = 2\mu$. 
The bosonic part of the lagrangian is thus a sum of three sinh-Gordon models.
This is exactly what one should expect from the fact that 
the bosonic subalgebra of $D(2,1;\A)$ is $A_1 \oplus A_1 \oplus A_1$
\cite{EM}. 
These three terms are nothing but the bosonic components of 
the three super-sinh Gordon lagrangians above.
We emphasize again that one of the sinh-Gordon models appears with an overall
minus sign, due to the indefinite signature of the 
invariant inner-product. 

Considering the remaining terms,
the fermionic part of the lagrangian is simply 
$$
L_{\rm ferm} = -{1\over 1+\A} L_{\rm free} (\psi_1) + L_{\rm
free}(\psi_2) 
+ {1\over\A}L_{\rm free}(\psi_3) 
$$ 
where we use the notation
$$L_{\rm free} (\psi) = i\psi_+\del_-\psi_++i\psi_-\del_+\psi_-$$
for the free lagrangian for a massless fermion.
Finally the interaction lagrangian is 
\bea 
L_{\rm int}  & = & i \mu \sum (\A_j \cdot \psi_+)(\A_j \cdot \psi_-)
\exp (\A_j \cdot \phi) \nn
&= & m ( \bar{\psi}_1\psi_1 + \bar{\psi}_2\psi_2 + 
\bar{\psi}_3\psi_3 ) 
\left ( 
\cosh\phi_1 \cosh\phi_2 \cosh\phi_3 -
\sinh\phi_1\sinh\phi_2\sinh\phi_3 \right ) \nn
&& + \, 2 m \bar{\psi}_1\psi_2 \left ( 
\sinh\phi_1\sinh\phi_2\cosh\phi_3 - 
\cosh\phi_1 \cosh\phi_2\sinh\phi_3 \right ) \nn
&& + \, 2 m \bar{\psi}_1\psi_3 \left ( 
\sinh\phi_1\cosh\phi_2\sinh\phi_3 - 
\cosh\phi_1\sinh\phi_2 \cosh\phi_3\right ) \nn
&& + \, 2 m \bar{\psi}_2\psi_3 \left ( 
\cosh\phi_1\sinh\phi_2\sinh\phi_3 - 
\sinh\phi_1\cosh\phi_2\cosh\phi_3\right ) 
\ena 
where the usual Lorentz-invariant inner-product for spinors is 
$\bar{\psi} \chi = i(\psi_+\chi_-+\chi_+\psi_-)$. 

\subsection{Twisted Reality Conditions}

Having written down the model, our next step is to discuss the
imposition of different reality conditions, analogous to our
discussion of (\ref{twist}) and (\ref{stwist}) in the introduction.
We repeat that our aim is to 
change the bosonic subsector from a sum of three
sinh-Gordon theories with a relative sign
between them (coming from the indefinite signature of the
inner-product 
on the target space $V$) to a combination of sine/sinh-Gordon
theories which all contribute with the same sign.

In \cite{JE} it was pointed out that any  
symmetry of order 2 of the Dynkin diagram for a Lie (super)algebra 
gives us a choice of reality conditions for the Toda fields consistent
with integrability. Specifically, if $\tau$ is the automorphism
acting on simple roots $\alpha_i$, we can impose the reality condition
\be\label{gentwist}
\alpha_i \cdot \Phi^* = \tau(\A_i) \cdot \Phi
\ee
where $^*$ denotes complex conjugation.
One can think of this as arising 
from the Toda theory based on the particular real form of the
algebra which is defined as the fixed-point set of the anti-linear 
conjugation map $\widehat{\tau}$, where 
$$
\widehat{\tau}( H_{\alpha_i} ) = H_{\tau (\alpha_i)} \, , \qquad
\widehat{\tau}( E_{\pm \alpha_i} ) = E_{\pm \tau (\alpha_i)}  \ .
$$
Taking $\tau$ to be the identity corresponds to the trivial symmetry
of the Dynkin diagram. Then the Toda fields
are all real, and the real form of the algebra is split, or 
maximally non-compact.
For non-trivial symmetries $\tau$ more interesting possibilities
arise, however.

Applying this to the case at hand,
the Dynkin diagram of $D(2,1;\A)^{(1)}$ has symmetry 
$\fl{Z}_2 \times \fl{Z}_2$
with three non-trivial 
elements\footnote{For $\alpha=1$, when $D(2,1;\A)$ 
reduces to $D(2,1)$, there 
is an additional symmetry corresponding to the permutation of the roots 
$\alpha_2$ and $\alpha_3$---it is not relevant now but will be
discussed again in section 4.} which act on the Dynkin diagram: 
$$
\hbox{\raise 30pt \hbox{(1):}\quad\hbox{\epsfxsize=70pt \epsfbox{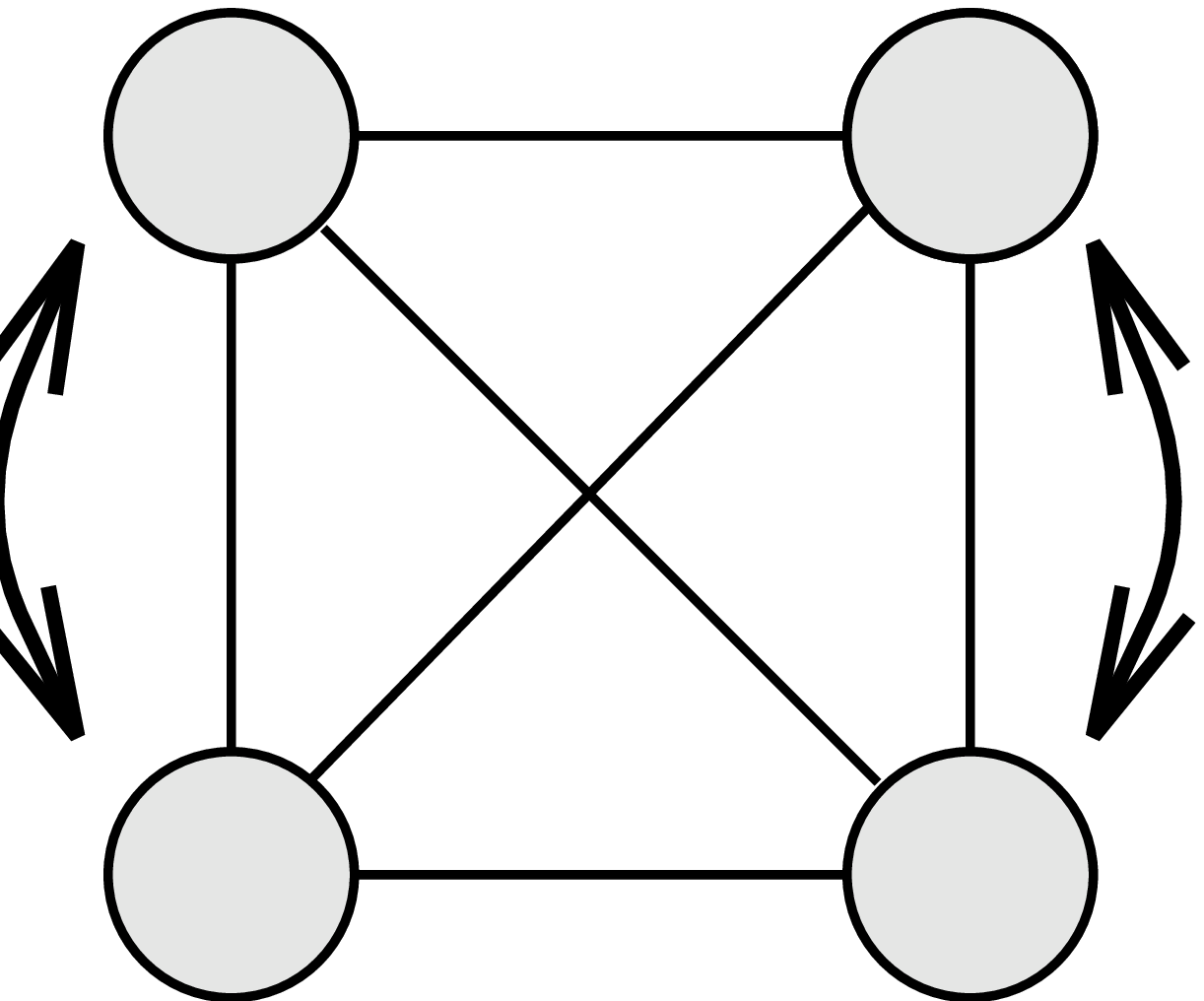}}}
\hspace{30pt}
\hbox{\raise 30pt \hbox{(2):}\quad\hbox{\epsfxsize=65pt \epsfbox{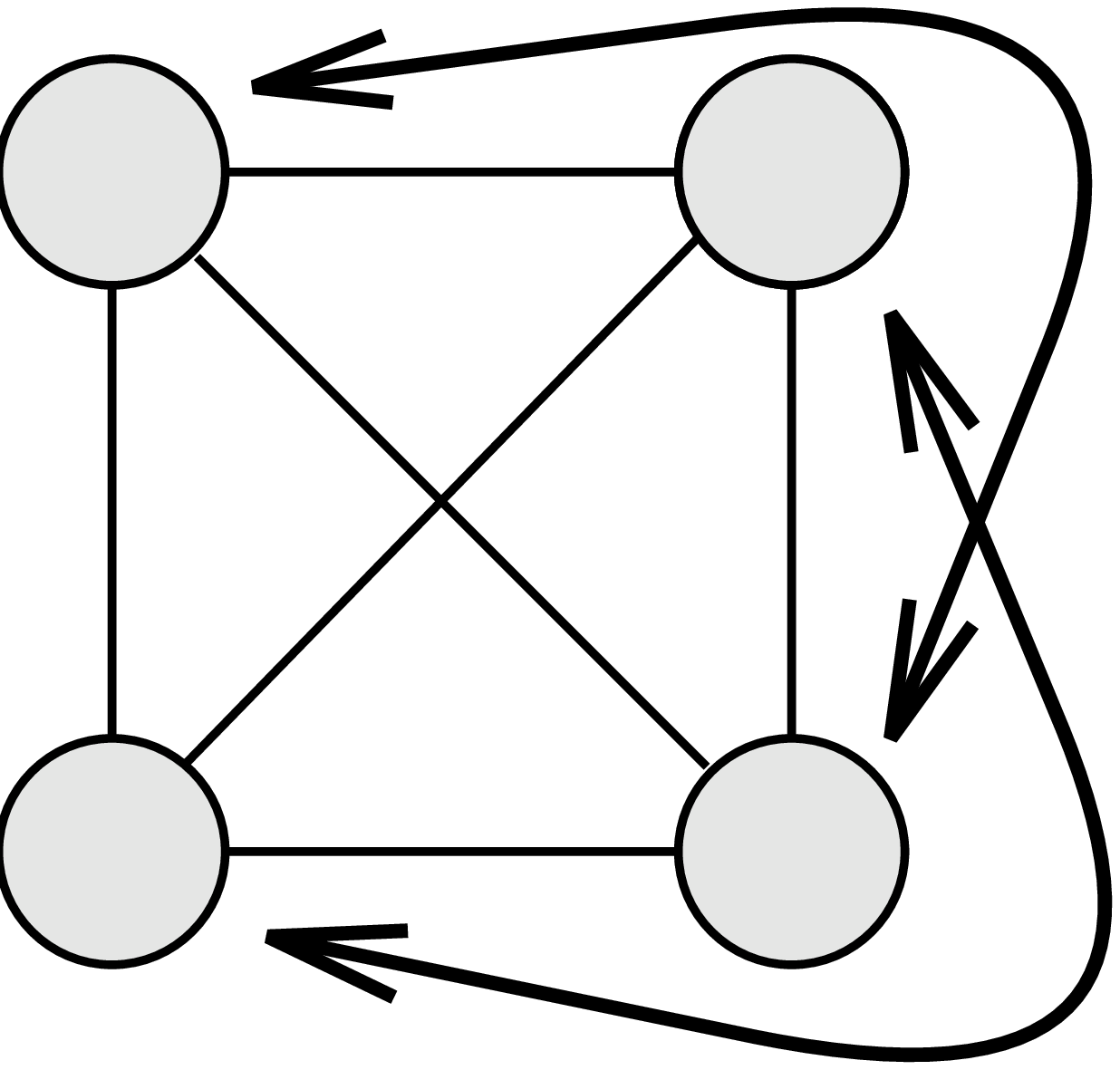}}}
\hspace{30pt}
\hbox{\raise 30pt \hbox{(3):}\quad\hbox{\epsfxsize=55pt \epsfbox{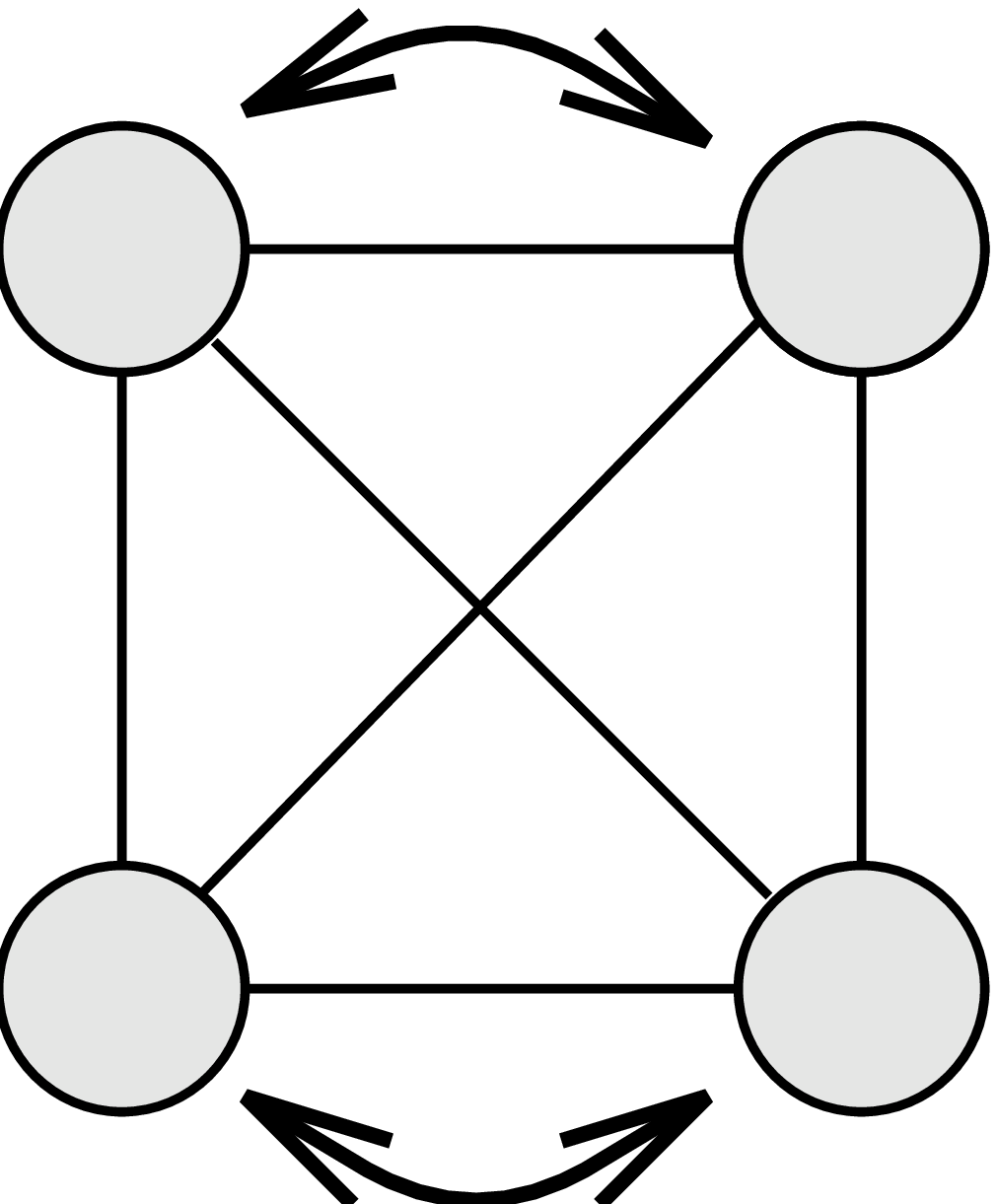}}}
$$ 
We have omitted the values of the bonds for clarity, but they are 
as in the Dynkin diagram (\ref{DD}). 
In terms of our orthogonal basis vectors the symmetries depicted 
above can be written: 
\be
\label{sym} 
\begin{array}{cccrccrccr}
(1) &  \e_1 & \rightarrow & \e_1, &  \e_2 & \rightarrow & 
-\e_2,& \e_3 & \rightarrow & -\e_3\,; \\
(2) & \e_1 & \rightarrow & 
-\e_1,& \e_2 & \rightarrow & \e_2,& \e_3 & \rightarrow & -\e_3
\,;\\ 
(3) & \e_1 & \rightarrow & -\e_1,& \e_2 & 
\rightarrow & -\e_2, & \e_3 & \rightarrow & \e_3
\,.
\end{array} 
\ee
The product of any two of these symmetries gives the third.
Naturally these are also symmetries of the lagrangian, as can be 
seen from equation (\ref{susy}).  

Taking as an example the symmetry denoted by (1), we can therefore 
consistently impose reality conditions 
\be\label{dtwist}
(\Phi_1)^*=\Phi_1; \qquad (\Phi_2)^*=-\Phi_2; \qquad 
(\Phi_3)^*=-\Phi_3\ee
i.e. we can replace $\Phi_2$ by $i\widehat{\Phi}_2$ and 
$\Phi_3$ by $i\widehat{\Phi}_3$ where $\widehat{\Phi}_j$ are real fields
(and correspondingly for the components $\phi_j$ and $\psi_{j\pm}$).  
It is clear from e.g. (\ref{susy}) that this does indeed give a real
lagrangian.
Equivalently, we can say that the space of fields $V$ is now the real span
of the vectors 
\be 
{\hat \e}_1 = \e_1; \qquad {\hat \e}_2 = i \e_2; \qquad {\hat \e}_3 = i \e_3
\ee
The inner-product on $V$ is clearly changed.

For each of the symmetries (1), (2), (3) 
we can introduce modified basis vectors 
for the target space $V$ as above.
The inner-product is diagonal in this basis, 
and it will be positive (negative) definite if 
all the squares of the bosonic roots are positive (negative). 
Examining each case, we find: 
$$
\begin{array}{crrrcl}
\mbox{symmetry} & 2 {\hat\e}_1^2 \mbox{~~~~}& 2{\hat \e}_2^2 & 2{\hat \e}_3^2 
& \mbox{~~~~} & \mbox{inner-product:} \\ 
(0) & -(1+\A) & 1 & \A && \mbox{indefinite} \\ 
(1) & -(1+\A) & -1 & -\A && \mbox{negative-definite for }\A>0 \\ 
(2) & (1+\A) & 1 & -\A && \mbox{positive-definite for }
-1 < \A < 0 \\ 
(3) & (1+\A) & -1 & \A && \mbox{negative-definite for }\A<-1 
\end{array}
$$ 
where we have denoted the trivial symmetry, corresponding to standard
reality conditions, by (0).

Given our earlier discussion about the isomorphic nature of algebras 
with parameters 
$\A$ related by the transformations (\ref{iso}), 
we can without loss of generality assume that
$\A$ is positive.
Having made this choice, we see that that 
the reality conditions (\ref{dtwist}) 
corresponding to the symmetry (1) are the unique choice which result 
in a lagrangian with definite signature. 
The theory with standard reality conditions corresponds to
the split real form $D(2,1;\A;0)$ of $D(2,1;\A)$, with bosonic 
subalgebra $sl(2,\fl{R})\oplus sl(2,\fl{R})\oplus sl(2,\fl{R})$, while the 
theory 
with twisted reality conditions corresponds to a Toda theory 
based on the real form $D(2,1;\A;1)$ with bosonic 
subalgebra $su(2)\oplus su(2)\oplus sl(2,\fl{R})$. 
(Note that only the definiteness of the signature is relevant; classically 
it is not important whether the signature is positive or negative, and when 
we consider the quantum theory we will simply choose the overall sign of the 
lagrangian such that the signature of the kinetic part is positive definite.) 

For ease of notation we shall now
drop the hats on the new real fields, simply replacing our original fields 
$\Phi_2 \rightarrow i\Phi_2$ and $\Phi_3 \rightarrow i\Phi_3$ 
and similarly for their components $\phi_2,\,\phi_3$ and $\psi_2,\,\psi_3$. 
In terms of these new real superfields the lagrangian is
\be
\label{twsusy}
{\cal L} = {-i\over 1+\A} D_+\Phi_1 D_-\Phi_1 - i D_+\Phi_2
D_-\Phi_2 - {i\over\A}D_+\Phi_3 D_-\Phi_3 + \mu U(\Phi) \ .
\ee
where 
\be
\label{twpot}
U(\Phi) = 4 ( \cosh \Phi_1 \cos \Phi_2 \cos \Phi_3 + \sinh \Phi_1 \sin
\Phi_2 \sin \Phi_3 )
\ee
The component lagrangian takes the form: 
\bea
\label{sin}
L_{\rm bos} & = & -{1\over 1+\A} \left (
\del_+ \phi_1 \del_- \phi_1 - 2\mu^2(1+\A)^2 \cosh 2\phi_1 \right ) \nn
&& \quad 
- \left ( \del_+ \phi_2 \del_- \phi_2 + 2\mu^2 \cos 2\phi_2 \right )\nn
&& \quad  
- {1\over\A} \left(
\del_+ \phi_3 \del_- \phi_3 + 2\mu^2\A^2 \cos 2\phi_3 \right ) \nn
& = & -{1\over 1+\A} L_{\rm sinh}(\phi_1;m(1+\A))-L_{\rm sin}(\phi_2;m)-
{1\over \A} L_{\rm sin}(\phi_3;m\A)
\ena
where the sine-Gordon lagrangian $L_{\rm sin}$ is defined in (\ref{bsin}) 
and, as before, $m=2 \mu$. 
The fermionic part of the lagrangian becomes
$$ 
L_{\rm ferm} = -{1\over 1+\A} L_{\rm free}(\psi_1) - L_{\rm
free}(\psi_2) - 
{1\over\A}L_{\rm free}(\psi_3),
$$ 
while the interaction lagrangian is 
\bea 
L_{\rm int}  & = & 
 m (\bar{\psi}_1\psi_1-\bar{\psi}_2\psi_2-\bar{\psi}_3\psi_3) \left ( 
\cosh\phi_1 \cos\phi_2 \cos\phi_3 + \sinh\phi_1\sin\phi_2\sin\phi_3
\right ) \nn
&& - \, 2m\bar{\psi}_1\psi_2 \left ( \sinh\phi_1\sin\phi_2\cos\phi_3 - 
\cosh\phi_1 \cos\phi_2\sin\phi_3 \right ) \nn
&& - \, 2m\bar{\psi}_1\psi_3 \left (\sinh\phi_1\cos\phi_2\sin\phi_3 - 
\cosh\phi_1\sin\phi_2 \cos\phi_3\right ) \nn
&& + \, 2m\bar{\psi}_2\psi_3 \left ( \cosh\phi_1\sin\phi_2\sin\phi_3 +
\sinh\phi_1\cos\phi_2\cos\phi_3\right ).
\ena 
We see that 
the lagrangian indeed has definite signature, as desired,
with the bosonic part being a sum of sinh-Gordon model and two
sine-Gordon models. 

\subsection{Similarities and Differences}

The main point is of course the profound difference between 
(\ref{sinh}), with indefinite kinetic energy and unbounded potential,
and (\ref{sin}), with definite kinetic energy and a potential bounded from
below. The quantization of our new theory with twisted reality
conditions is (at least
conceptually) straightforward; whereas the meaning which can be
ascribed to the quantization of the original model with standard
reality conditions is delicate to say
the least. 

Nevertheless, there are also some strong similarities.
The classical masses of bosonic and fermionic particles 
are known to be independent of the the choice of reality conditions \cite{JE}.
For the bosons they can be read off from 
(\ref{sinh}) or (\ref{sin}): 
$$ 
m_1^2 = m^2 (1+\A)^2;\quad m_2^2 = m^2;\quad
m_3^2 = m^2 \A^2 
$$ 
The masses of the fermions can be found by
expanding the interaction lagrangian to second order in the fields,
the results being 
$$ 
m_1 = m (1+\A); \quad m_2 = - m; \quad m_3 = - m\A 
$$ 
consistent with supersymmetry.
We note that there is a relative sign between the masses for the
fermions. However, as
in the case of bosons, it is the square of the mass that has a physical 
interpretation; the Dirac equation with a negative mass has the same space 
of solutions as the Dirac equation with positive mass, but with solutions with 
negative and positive energy interchanged. 
Notice also that the masses of the various particles in the theory are given 
by the entries in the Cartan matrix of $D(2,1;\A)$, and that the particles can 
therefore be directly assigned to the three bonds of the Dynkin diagram of 
$D(2,1;\A)$; this should be compared to the Toda theory for a bosonic Lie 
algebra 
$\cg$, where the particles 
in the $\cg^{(1)}$ Toda theory can be assigned to the {\em nodes} of 
the Dynkin diagram of $\cg$.  

In the last section of this paper we shall show that the integrability 
of our model extends to the quantum theory, giving hope that its
S-matrix could be determined exactly.
Any candidate S-matrix must be compared with perturbative scattering
calculations, the starting point for which are the
classical three-point functions.
Note that the symmetries (\ref{sym}) of the lagrangian restrict the
non-zero three-point couplings $c^{ijk}$ to be $c^{123}$. 
Expanding the lagrangian with 
normalized fields to third order, we find the explicit Yukawa interactions: 
$$
L_{\rm yuk} =  -4\mu\sqrt{\A(1+\A)}
(\bar{\psi}_1\psi_2\phi_3+\bar{\psi}_1\psi_3\phi_2+\bar{\psi}_2\psi_3\phi_1)
$$ 

The last important difference to be emphasized concerns the non-trivial 
vacuum structure of the sine-Gordon terms appearing in (\ref{sin}), as
against the unique vacua of the sinh-Gordon theories in (\ref{sinh}).
The sine-Gordon model with lagrangian (\ref{bsin})
has an infinite
set of classical vacua at $\phi=2p\pi$ and hence 
soliton and anti-soliton solutions of the 
classical equations of motion which interpolate between these, 
with $\phi$ tending to different vacuum values as $x \rightarrow
\pm\infty$.
Explicitly these solitons are given (in their rest-frames) by: 
$$
\phi^\pm = 4 \tan^{-1}[\exp(\pm{m\over 2}(x^+-x^-))]
$$
When quantizing the theory (see e.g.~\cite{Ra}) one finds the 
surprising result that the solitons and 
anti-solitons are in fact more fundamental than 
the particles corresponding to the field $\phi$, since the latter 
can be identified with the breather
bound states of a soliton and an anti-soliton.

In the supersymmetric sine-Gordon theory, 
the classical bosonic solitons are expected on general grounds to give
rise to a pair of quantum states with 
fermion numbers $\pm \hf$ which form a doublet under supersymmetry.         
The precise way in which these structures appear in the 
semi-classical spectrum of the $D(2,1;\A)^{(1)}$ model is 
of crucial importance for determining its S-matrix.
At the very least we expect
to find two species of solitons and anti-solitons 
arising from the 
two sine-Gordon theories in (\ref{sin}) 
as well as the fundamental bosons and fermions of the sinh-Gordon
sector.
A more detailed analysis of the 
particle spectrum in the quantum theory 
is work for the future, however. 

\section{Other Lie Superalgebras}
\label{other}

In this section we take a slight detour in order to consider 
some related Toda models.
In our previous paper \cite{EM} we showed that only a very few 
supersymmetric affine Toda theories exist which share with 
the $D(2,1;\A)^{(1)}$ family the important property that there are
consistent reality conditions for the fields which render the kinetic energy 
positive-definite. The other Lie
superalgebras in question are: 
$$
C(2)^{(2)}\simeq A(1,0)^{(2)},\qquad B(1,1)^{(1)},\qquad 
A(1,1)^{(1)}
$$
We shall describe how all these can be obtained from the 
$D(2,1;\A)^{(1)}$ family in various limits. 
This material is
not essential for understanding the $D(2,1;\A)^{(1)}$ example itself,
but it does underscore its importance.

\subsection{The algebras $C(2)^{(2)}$ and $B(1,1)^{(1)}$}

To obtain these algebras from $D(2,1;\A)^{(1)}$ 
we use the procedure of 
{\em folding}, as discussed in e.g.~\cite{FSS}. 
Given a symmetry $\tau$ of order 2 of any Dynkin diagram 
of a Lie superalgebra, the folded algebra 
is defined to have a Dynkin diagram with simple roots 
$\hf(\A+\tau(\A))$; 
this corresponds exactly to the 
$\tau$-invariant subalgebra. 

In section \ref{toda} we discussed the symmetries of the Dynkin diagram 
of $D(2,1; \A)^{(1)}$ in some detail in connection with possible reality
conditions for the Toda fields. 
Taking the symmetry (1) in (\ref{sym}), for instance, and applying
the folding construction we
obtain the Dynkin diagram of $C(2)^{(2)}$:\\[6pt]
\hbox{\hspace{100pt}\hbox{\epsfxsize=70pt \epsfbox{sym1.eps}}\hspace{25pt} 
\raise 28pt \hbox{$\longrightarrow$}\hspace{20pt}
\raise 25pt \hbox{\epsfxsize=60pt \epsfbox{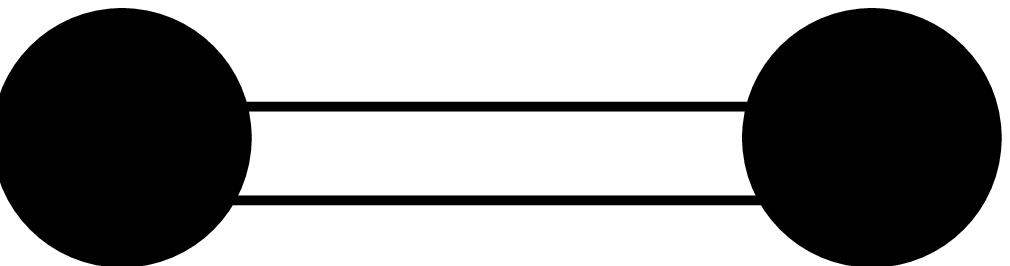}}}\\[5pt]
The folded algebra is generated by simple roots 
$\hf(\A_2+\A_3)$ and $\hf(\A_0+\A_1) = - \hf(\A_2+\A_3)$. 
In terms of the fields $\Phi_i$ in the Toda model, this 
corresponds to setting $\Phi_2=\Phi_3=0$ and we see 
immediately that the result 
is just the $N=1$ sinh-Gordon model given previously in (\ref{ssg1}).
Folding by either of the other symmetries (2) or (3) gives an 
isomorphic result, corresponding to the other sinh-Gordon sub-theories
in (\ref{ssg2}) and (\ref{ssg3}).

Notice that the use of a symmetry $\tau$ in the folding construction 
in entirely different from its use in defining 
twisted reality conditions (\ref{gentwist}).
Actually, it make sense to consider 
folding by one symmetry and twisting by another simultaneously, provided that 
the latter operation is based on a symmetry of the
original diagram which survives after folding is carried out.
Thus, in the example above, either of the symmetries (2) or (3) 
reduces to the unique reflection symmetry of the folded diagram
for $C(2)^{(2)}$.
Twisted reality conditions in the original theory
will then descend to the Toda theory based on the folded algebra in an
obvious manner. The effect 
on the super-sinh-Gordon system is of course exactly the change 
(\ref{stwist}) which gives the super-sine-Gordon system.

To discuss the next example we first set 
$\A =1$ to obtain $D(2,1)^{(1)}$.
For this special value of the parameter $\A$ there is an additional
symmetry of the Dynkin diagram beyond those we discussed earlier in
(\ref{sym}).
When $\A=1$ it is clearly possible to exchange the simple roots 
$\alpha_2 \leftrightarrow \alpha_3$ while keeping 
$\alpha_0$ and $\alpha_1$ fixed, or equivalently to take 
$\e_2 \leftrightarrow \e_3$ in terms of our orthogonal basis vectors.
One may wonder whether this symmetry gives rise to
another interesting choice of reality conditions in the $D(2,1)^{(1)}$
Toda theory itself, but in fact it does not 
lead to a model with definite kinetic energy, which is the reason we
omitted it from our earlier discussion.

By folding the Dynkin diagram of $D(2,1)^{(1)}$ using this new symmetry we 
obtain the 
Dynkin diagram of $B(1,1)^{(1)}$, as shown\footnote{ 
When considering $\A=1$ we revert to the standard way of drawing the Dynkin 
diagram, so the number of bonds between roots $\A_i$ and $\A_j$ is  
max$(|c_{ij}|,|c_{ji}|)$.}: \\[6pt]
\hbox{\hspace{100pt}\hbox{\epsfxsize=70pt \epsfbox{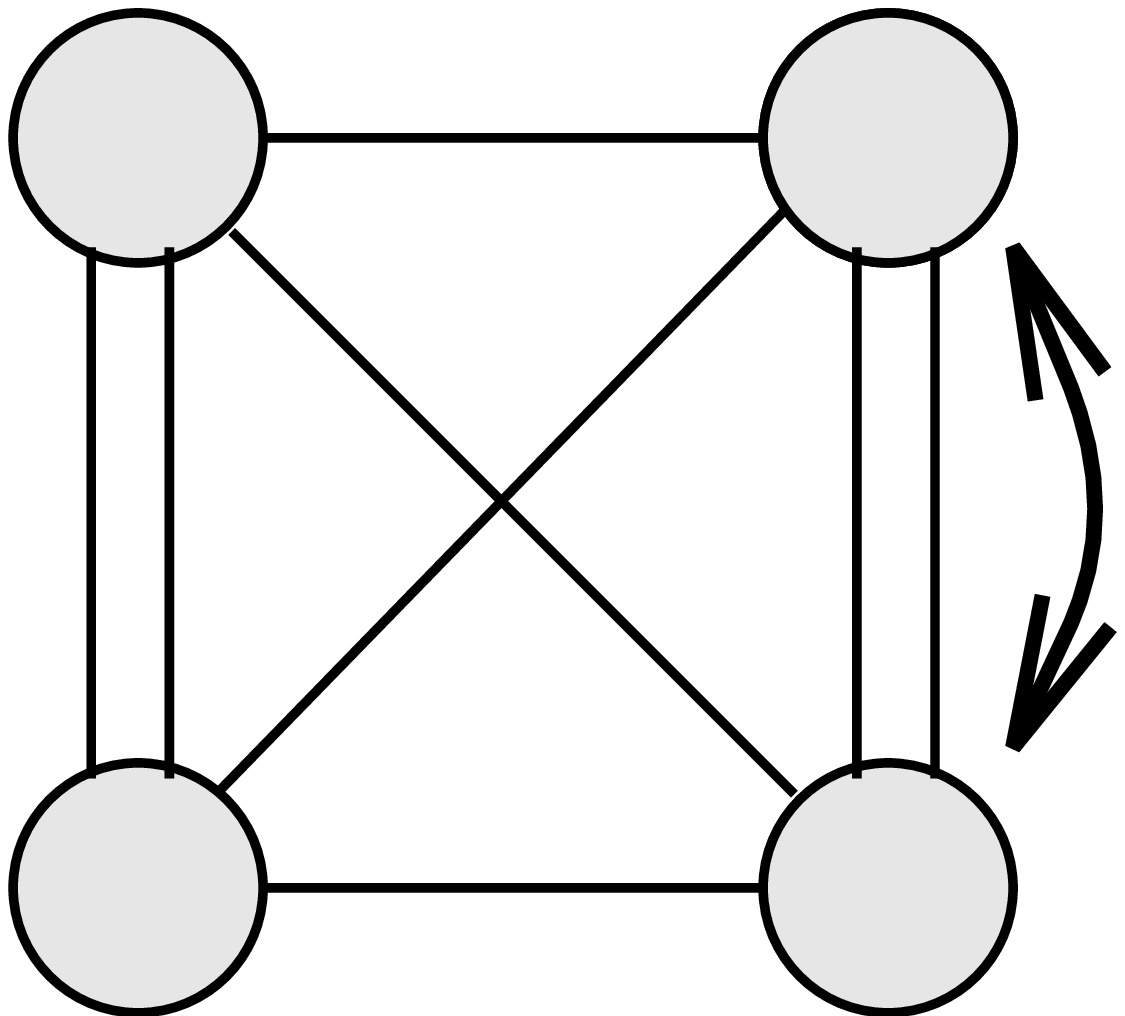}}\hspace{25pt} 
\raise 28pt \hbox{$\longrightarrow$} \hspace{20pt} 
\raise 5pt \hbox{\epsfxsize=40pt \epsfbox{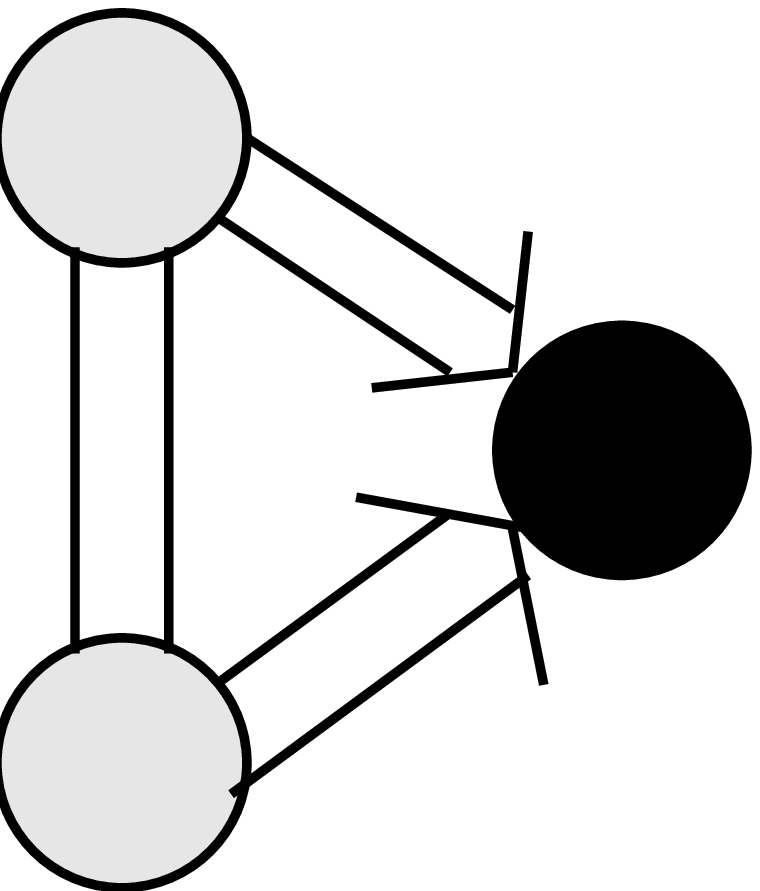}}}\\[5pt]
The new algebra has simple roots 
$\A_0,\,\A_1$ and
$\hf(\A_2+\A_3)$. In terms of the Toda fields $\Phi_i$ this corresponds to 
the identification $\Phi_2=\Phi_3$ so that the superspace lagrangian
is
\be
{\cal L} = - {i \over 2} D_+ \Phi_1 D_- \Phi_1 + 2 i D_+ \Phi_2 D_-
\Phi_2 + 2m ( \cosh \Phi_1 \cosh^2 \Phi_2 - \sinh\Phi_1 \sinh^2
\Phi_2 )
\ee
Once again we have the possibility of taking twisted reality conditions 
corresponding to the symmetry (1) of our original diagram, which
descends to the folded diagram by exchanging the two grey nodes.
This amounts to replacing $\Phi_2 \rightarrow i \Phi_2$ 
to give 
\be
{\cal L} = {i \over 2} D_+ \Phi_1 D_- \Phi_1 + 2 i D_+ \Phi_2 D_-
\Phi_2 + 2m ( \cosh \Phi_1 \cos^2 \Phi_2 + \sinh\Phi_1 \sin^2
\Phi_2 )
\ee
The component content of these theories 
(which were briefly discussed in
\cite{EM}) can readily be examined in detail by making the
appropriate identifications in the general expressions given earlier 
in section 3.
It was shown in 
\cite{QI2} that there are 
higher-spin quantum conserved currents in this model 
(assuming standard reality conditions). 

\subsection{The algebra $A(1,1)^{(1)}$}

The case of $A(1,1)^{(1)}$ differs from the other algebras 
we have considered in a number of respects. 
One reason is that this 
Toda model generally allows for two independent mass parameters, 
or equivalently one mass parameter together with an independent 
dimensionless coupling.
Another reason is that the algebra arises as a kind of singular limit 
of $D(2,1;\A)^{(1)}$ rather then by through the folding construction.

As explained in section \ref{d21}, the parameter $\A$ in $D(2,1;\A)^{(1)}$
takes values in $\fl{C} \setminus \{-1,0\}$. If we take 
$\A\in\{-1,0\}$ various properties like the dimension and rank of the
algebra change
discontinuously, but we still obtain a sensible Lie superalgebra nevertheless.
On choosing either of these values, 
the Cartan matrix of $D(2,1;\A)^{(1)}$ reduces to the
Cartan matrix of $A(1,1)^{(1)}$:
$$
\hbox{\raise 20pt 
\hbox{$
\left ( \begin{array}{cccc} 
0 & -1 & 1  & 0 \\ -1 & 0 & 0 & 1 \\ 1 & 0 & 0 & -1 \\
0 & 1 & -1 & 0
\end{array} \right )$} \hspace{15mm}
\hbox{\epsfxsize=55pt \epsfbox{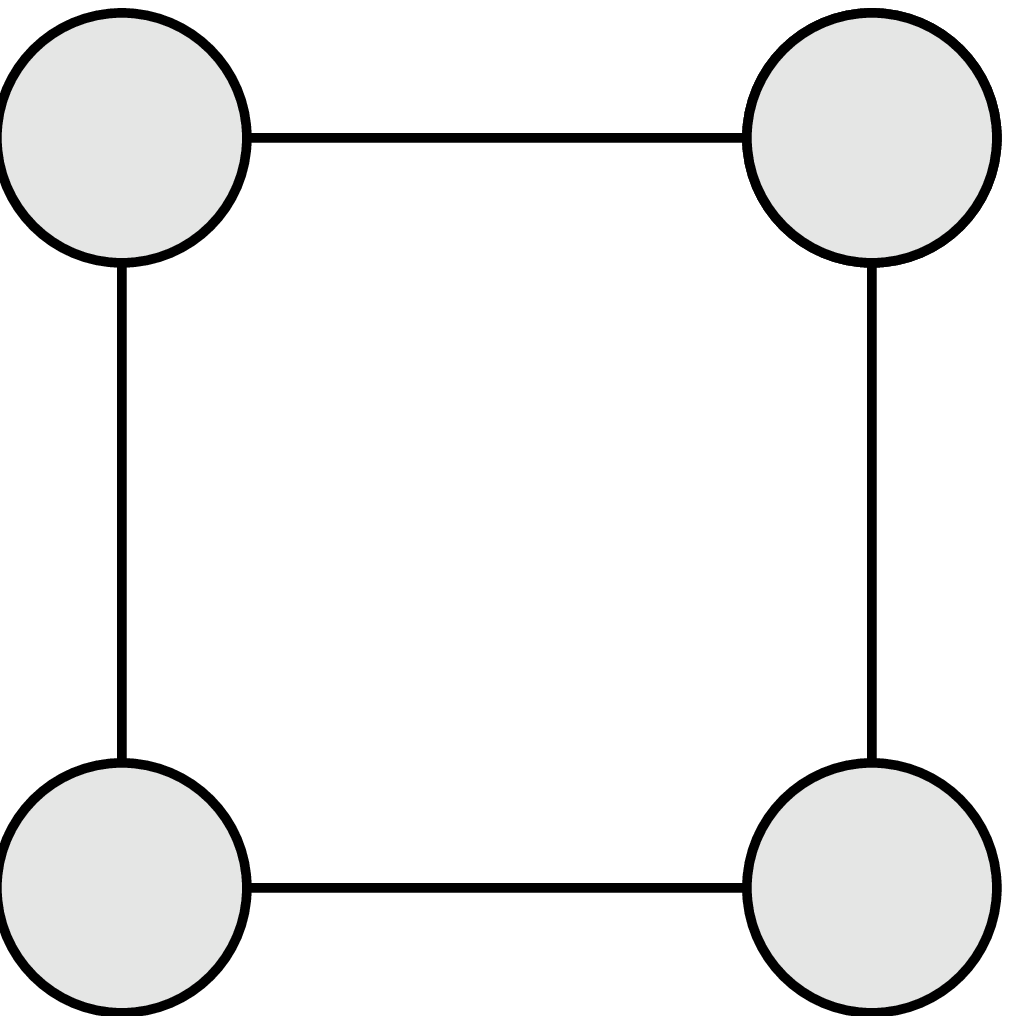} }}
$$
where the ordering of rows and columns corresponds to a `non-standard' 
ordering of nodes in the Dynkin diagram.
In terms of the simple roots of $D(2,1;\A)$,
taking $\A=0$ implies the identification $\A_2=-\A_1$.

The corresponding Toda theory has fields $\Phi_1$ and $\Phi_2$ which
can
be thought of as descending directly from the $D(2,1;\A)^{(1)}$ model;
the new relation amongst the simple roots implies that we must set 
$\Phi_3=0$. The lagrangian for the $A(1,1)^{(1)}$ model is 
$$
{\cal L} = -iD_+\Phi_1 D_-\Phi_1 + i D_+\Phi_2 D_-\Phi_2 
+\mu \cosh(\Phi_1+\Phi_2) + \nu \cosh(\Phi_1-\Phi_2)
$$
which is slightly more general than those we have 
considered previously because it involves 
two mass parameters,
$\mu$ and $\nu$. 
The appearance of a second mass parameter is due to the fact that 
the Cartan matrix of $A(1,1)^{(1)}$ has {\em two} null eigenvectors
or, in other words, because there are two linear relations between the four
simple roots of $A(1,1)^{(1)}$. 

With standard reality conditions, the superfields $\Phi_i$ 
are real and so are the constants $\mu$ and $\nu$.
What is of more interest to us, however, is that we can impose 
twisted reality conditions to obtain a
positive-definite action by using the symmetry (2) in (\ref{sym}) 
of the original diagram, which amounts to $\e_1 \rightarrow - \e_1$
in this degenerate limit.
This effectively means that we replace 
$\Phi_1 \rightarrow i \Phi_1$. In addition the mass
parameters are now allowed to be complex too, provided 
they are related by $\mu = \nu^*$. 
It is convenient to set $\mu = m e^{i \gamma}$ and 
$\nu = m e^{-i \gamma}$ and then the lagrangian becomes 
$$
{\cal L} = iD_+\Phi_1 D_-\Phi_1 + i D_+\Phi_2 D_-\Phi_2 
+2 m \cos \gamma \cos\Phi_1 \cosh\Phi_2 -  2 m \sin \gamma
\sin\Phi_1\sinh\Phi_2 
$$
which is clearly real.
In addition to the mass-scale $m$ it involves the 
real dimensionless parameter $\gamma$.

On the face of it, it seems that we have found another one-parameter
family of theories with positive-definite kinetic energy, similar to
our $D(2,1;\A)^{(1)}$ family, though with a slightly different origin 
for the continuous parameter.
But in fact the models written above 
are not really distinct at all because 
the value of $\gamma$ can be changed arbitrarily by a suitable 
field re-definition. 
This behaviour also turns out to be related to another very special 
property, namely the 
occurrence of an extra supersymmetry.

To explain how all this comes about, let us 
examine the component content of the
theory. On expanding the superfields and 
eliminating the auxiliary fields in the usual way we find the 
component lagrangian: 
$$
L = L_{\rm bos} + L_{\rm ferm} + L_{\rm int} 
$$
with 
\beano
L_{\rm bos} & = & \del_+\phi_1\del_-\phi_1 + \hf {m^2 } \cos 2 \phi_1
+ \del_+\phi_2\del_-\phi_2 - \hf {m^2 } \cosh 2 \phi_2\\
L_{\rm ferm} & = & L_{\rm free}\,(\psi_1) + L_{\rm free}\,(\psi_2)
\enano
where $L_{\rm free}$ denotes the free fermion lagrangian as before, and 
\beano
L_{\rm int} & = & m \cos \gamma \left [ 
(\bar{\psi_1}\psi_1 - \bar{\psi_2}\psi_2)\cos\phi_1\cosh\phi_2-
2\bar{\psi_1}\psi_2\sin\phi_1\sinh\phi_2 \right ] \\
&& - m \sin \gamma \left [ 
(\bar{\psi_1}\psi_1 - \bar{\psi_2}\psi_2)\sin\phi_1\sinh\phi_2 +
2\bar{\psi_1}\psi_2\cos\phi_1\cosh\phi_2 \right ]
\enano
In the bosonic sector we clearly have sine-Gordon and sinh-Gordon 
models, each with mass $m$. 
The lagrangian is supersymmetric, by construction, for any value of
the dimensionless parameter $\gamma$. The fermions must therefore
be degenerate in mass with the bosons, which can be confirmed by
diagonalizing their mass matrix.

Now we claimed above that different values of $\gamma$ do not really 
give distinct theories, and the key to understanding this is that 
the value of $\gamma$ can be changed by a rotation of the fermions.
Specifically, the transformation
$$
\left [ \begin{array}{c}\psi_1\\ \psi_2 \end{array} \right ]
\rightarrow 
\left [ \begin{array}{rr} \cos\theta&\sin\theta\\
-\sin\theta&\cos\theta\end{array}\right ] 
\left [ \begin{array}{c}{\psi_1}\\ {\psi_2}\end{array} \right ] 
$$
on the fermions is actually indistinguishable from the shift
$$
\gamma \rightarrow \gamma + 2 \theta
$$
as can be easily checked from the formula for 
the interaction lagrangian (the other terms in the lagrangian 
are manifestly unchanged).
This means of course that the value of $\gamma$ has no physical
significance, since it can be changed at will by this field
re-definition. It also means incidentally that we can choose $\theta$
so as to set $\gamma= 0$ and thereby read off the fermion masses $\pm m$,
confirming our previous statement.

The final part of the story is the link with $N=2$
supersymmetry. 
Although the value of $\gamma$ is unimportant in itself, a change in
$\gamma$ can only be produced by a rotation of the fermions.
We stressed above that the lagrangian is supersymmetric for any
value of $\gamma$, but implicit in this statement is the fact that the
bosons $\phi_j$ have the fermions $\psi_j$ appearing in 
$L_{\rm int}$ as their superpartners. 
If we change $\gamma$ and rotate the fermions appropriately to compensate, 
then the model does not change, but in doing so we obtain new 
supersymmetry transformations relating the bosons and fermions, so that
we actually have a continuous family of 
supersymmetries $Q_\gamma$. This is precisely what we mean by $N=2$
supersymmetry, although it is more usually characterized 
by saying that there are two 
independent supercharges, in our present notation these 
correspond simply to $Q_0$ and $Q_{\pi/2}$, for example.

To summarize: The affine Toda theory defined by $A(1,1)^{(1)}$ 
allows two independent real mass parameters in general.
But for the choice of reality conditions leading to a
positive-definite theory,
the ratio of these parameters 
has no physical significance and can be removed by a field
re-definition.
The resulting model is the $N=2$ super sine/sinh-Gordon theory, 
which has a single real mass parameter.
The disappearance of the additional 
dimensionless parameter is directly linked 
to the fact that we have a continuous set of supersymmetries,
i.e.~$N=2$ rather than $N=1$.

These conclusions can also be expressed very naturally using the 
language of $N=2$ superspace,
in which the rotation of the fermions discussed above becomes an
$R$-transformation on a complex chiral $N=2$ superfield containing
both $\Phi_1$ and $\Phi_2$. The $N=2$ superpotential term
is not invariant under such a transformation, but changes by a phase.
This implies that its coefficient, which is ultimately identified with 
the mass, can always be chosen to be real.

The remarks above are easily generalized to the whole family 
of $A(n,n)^{(1)}$ affine Toda theories and they provide a new way
of understanding 
why this family alone amongst $N=1$ affine Toda theories actually
admits $N=2$ supersymmetry (see \cite{EH}). It is only the 
$A(1,1)^{(1)}$ example for which the lagrangian has positive kinetic
energy, however.  

\section{Classical Conserved Currents} 
\label{ccons}

We now return to our analysis of the $D(2,1;\A)^{(1)}$ Toda model and
to the issue of its classical conserved currents.

\subsection{Preliminaries and Generalities}

The Lax pair construction mentioned earlier 
gives a systematic way of finding conserved quantities, thereby
ensuring the integrability of any Toda model.
In particular cases, however, we can also take a less sophisticated
but more direct approach to the construction of conserved quantities.
Let us discuss this first in the bosonic situation.

Conformal Toda models have 
`holomorphic' conservation laws of the form
\be \label{bconf}
\del_- J = 0
\ee
with the quantities $J$ generally forming a $\cw$-algebra.
The simplest case is the Liouville theory, based on the algebra
$sl(2;\fl{R})$,
with lagrangian 
\be \label{bliou}
L = \del_+ \phi \, \del_- \phi  - \mu \exp (2 \phi) \, .
\ee
Conservation of the (traceless) energy-momentum tensor 
corresponds to the fact that the following component 
is holomorphic: 
\be 
T = \del_+ \phi \, \del_+ \phi - \del_+^2 \phi
\ee 
We then see immediately 
that (\ref{bconf}) holds with $J=T^n$ for any integer $n$.

In massive Toda theories we have instead conservation laws
of the form
\be
\label{bcons}
\del_- J + \del_+ \bar J = 0
\ee
with both light-cone components of the current non-zero.
In passing from a conformally-invariant theory based on 
$\cg$ to a massive theory based on $\cg^{(1)}$, we may seek to
generalize a holomorphic conservation law by taking the same 
expression for $J$.
Since the equations of motion are modified in the massive theory (by the
addition of a term in the potential corresponding to the affine root) 
$J$ will no longer be
holomorphic, but it can often be made to satisfy (\ref{bcons})
for a suitable choice of $\bar J$.
The simplest example is again $\cg = sl (2;\fl{R})$, which 
entails passing from the
Liouville theory above to the sinh-Gordon model (\ref{bsinh}). It is easy to
check directly using the equations of motion from (\ref{bsinh}) that 
(\ref{bcons}) holds with
\be
J = T \, ; \qquad \bar J = - {m^2 \over 2} 
\exp (-2\phi) 
\ee
and 
\be
J = T^2 \, ; \qquad \bar J = - m^2 (\del_+ \phi)^2 \exp (- 2\phi)
\ee
Infinitely many higher-spin currents of this type could be 
found in a similar fashion.  
Notice that $\bar J$ always involves the 
term in the potential corresponding to the affine
root.

Much of this extends easily to superspace. 
The simplest superconformal Toda theory (the super-Liouville model)
has lagrangian
\be 
{\cal L} = \frac{i}2 D_+ \Phi D_- \Phi -  \mu \exp \Phi
\ee 
with conservation laws of the form
\be \label{sconf}
D_- J = 0 \ .
\ee
In particular, we find 
a super-holomorphic spin-3/2 super-energy-momentum tensor
\be 
T = D_+ \Phi D_+^2 \Phi - D^3_+ \Phi \ .
\ee
Since this is fermionic, we cannot take powers of it, but we can obtain
additional holomorphic currents through other combinations, 
such as the spin-7/2 current 
$T D_+ T$.

The affine extension of this model is precisely the 
super sinh-Gordon model (\ref{ssinh}).
Once again, the conformal conservation laws can be promoted to the
massive theory in the form
\be \label{scons}
D_+ J = D_- \bar J
\ee
by choosing $\bar J$ in a suitable way. 
Using the equations of motion from (\ref{ssinh}) 
it is straightforward to check that 
(\ref{scons}) holds with 
\be\label{sgem}
J = T \, ; \qquad 
\bar J =  {m \over 2} i D_+ (\exp (-\Phi)) 
\ee
and also with
\be\label{sghs}
J = T D_+ T  \, ; \qquad \bar 
J = {m \over 2} i ( D_+ \Phi (D_+^2 \Phi)^2 - 
D_+^2 \Phi D_+^3 \Phi ) \exp (-\Phi) \ .
\ee

To recap: we have a general way of searching for
conserved currents in the affine theory based on $\cg^{(1)}$ if we
have some knowledge of the conformal currents in the theory based on
$\cg$. This applies in either the bosonic or supersymmetric regimes.

One new feature does arise in the superspace situation, however.
We are well-accustomed in conformal field theory to the consideration
of superspace conservation equations (\ref{sconf}) which contain 
pairs of holomorphic currents related by supersymmetry.
Thus we know that the spin-3/2 superfield $T$ in the example above 
will contain a pair of conventional
fields with spins $({3 \over 2},2)$ 
which are the generators of supersymmetry and 
translations. Similarly, the spin-7/2 field will contain component
conserved quantities with spins $({7\over 2},4)$.
It is tempting to expect similar behaviour for conservation laws of
the form (\ref{scons}) but the following detailed analysis shows that this is
unfounded in general.\footnote{These considerations also arose
recently in a rather different context \cite{EHM}.}

Taking a general component expansion of the
current superfields:
\bea
J & = & \A + \theta^+ j + \theta^- h + i \theta^+ \theta^- \beta \nn
\bar J & = & \bar \A + \theta^- \bar \jmath + \theta^+ \bar h 
+ i \theta^- \theta^+ \bar \beta
\ena
we can examine the content of the equation (\ref{scons}) explicitly, and
we find
\bea
\del_- j + \del_+ \bar \jmath = 0 \label{jj} \\
\del_- \alpha = \bar \beta \label{ab} \\
\del_+ \bar \alpha = \beta \\
h = \bar h
\ena
(we assume for definiteness that $J$ and $\bar J$ are fermionic rather
than bosonic, then the component fields defined above are all real).
One of these equations is indeed a conservation equation for 
the bosonic current with components $(j, \bar \jmath)$. 
But the other equations give no additional conservation laws in the
absence
of any extra information about the structure of the currents.

An additional conservation law does arises, however, if
\be\label{spart}
\bar J = -i D_+ K
\ee
for some superfield $K$, since this implies 
\be
\bar \beta = - \del_+ \omega , \quad \bar h = - \del_+ k
\qquad {\rm where} \qquad
K = k + i \theta^+ \bar \alpha + i \theta^- \omega + i \theta^+
\theta^- \bar \jmath
\ee
and then (\ref{ab}) becomes a conservation equation 
\be\label{ao} 
\del_- \alpha + \del_+ \omega = 0
\ee
A simple calculation
reveals that the fermionic charge constructed from the current 
$(\alpha , \omega)$ and the
bosonic charge constructed from $(j, \bar \jmath)$ 
are indeed related by supersymmetry.
It is also simple to show that the condition on the superfield
(\ref{spart}) 
is the only way in which we can obtain an additional conservation law
of this type. (Assuming (\ref{ao}) and examining its change under a 
supersymmetry transformation restricts the other fields in ways which 
amount exactly to (\ref{spart}).)

The conformal case is a particularly simple way of satisfying 
(\ref{spart}) with 
$\bar J = K = 0$. A less trivial example of how this
condition can be met is provided by the conservation
of the super-energy-momentum tensor. In a supersymmetric theory it is
always possible to define a super-energy-momentum tensor so that
(\ref{spart}) is fulfilled.
It is easy to see this for models of the sort discussed in this paper, 
in which scalar superfields $\Phi$ interact via a lagrangian of the
general form
\be 
{\cal L} = \frac{i}2 D_+ \Phi \cdot D_- \Phi - V(\Phi)
\ee
where $V$ is some potential.
It follows from the equations of motion that 
\be 
D_- ( D_+ \Phi \cdot D_+^2 \Phi ) + i D^2_+ V(\Phi) = 0 
\ee 
(and similarly with $\pm$ exchanged).
Comparing with (\ref{scons}) we see
that this superspace conservation equation 
clearly involves a current fulfilling the additional 
property (\ref{spart}). 
This is only to be expected, since energy-momentum certainly should
come together with a conserved superpartner, namely 
the supersymmetry charge itself.

A special case of this analysis is provided by super-energy-momentum
conservation in the sinh-Gordon model (\ref{sgem}).
Notice however that the higher-spin conserved current (\ref{sghs})
does not meet the condition (\ref{spart}).
From the analysis above we know that it will produce a bosonic
conserved current of spin 4, but we do not expect this to have a
superpartner. 

\subsection{The $D(2,1;\A)$ models}

The \cw-algebra of conserved currents in the conformal Toda theory 
constructed from $D(2,1;\A)$ 
contains two generators of (super)spin ${3\over 2}$, and one generator of spin 
2, see e.g. \cite{SHR}. 
Explicit expressions for the currents 
(with standard reality conditions) are:
\bea
T & = & -{1\over 1+\A} ( D_+ \Phi_1 D^2_+\Phi_1 - D^3_+ \Phi_1 ) 
+ (D_+ \Phi_2 D^2_+\Phi_2 - D_+^3 \Phi_2 )
+ {1\over \A} ( D_+ \Phi_3 D^2_+\Phi_3 - D_+^3 \Phi_3) \nn
F & = & - {\A\over 1+\A}(D_+ \Phi_1 D_+^2 \Phi_1 - D_+^3 \Phi_1) + 
{1+\A\over \A}(D_+ \Phi_3 D_+^2 \Phi_3 - D^3_+ \Phi_3) \nn
&& - \, 
D_+ \left ( D_+\Phi_1D_+\Phi_2 + D_+\Phi_3D_+\Phi_1 + D_+\Phi_2D_+\Phi_3 
\right ) + 2 D_+\Phi_1D_+\Phi_2 D_+\Phi_3 \nn
W & = & D_+^4 \Phi_2 + {1\over \A} D_+^3\Phi_1 D_+\Phi_3 + 
D_+^3 \Phi_2 
\left({1\over 1+\A} D_+\Phi_1 - D_+\Phi_2 - {1\over\A}D_+\Phi_3 \right) \nn
&&
- \, {1\over 1+\A}D_+^3\Phi_3D_+\Phi_1 - (D_+^2\Phi_2)^2 \nn
& & - \, D_+^2(\Phi_1-\Phi_3) 
\left ({1\over 1+\A}D_+\Phi_1D_+\Phi_2 + {1\over \A(1+\A)}D_+\Phi_1D_+\Phi_3 - 
{1\over\A}D_+\Phi_2D_+\Phi_3 \right ) \nn
&& +  \, D_+^2\Phi_2 
\left ( {1\over 1+\A}D_+\Phi_1D_+\Phi_2 - 
{1+2\A \over \A(1+\A)}D_+\Phi_1D_+\Phi_3 + {1\over\A}D_+\Phi_2D_+\Phi_3 
\right ) 
\ena 
The first of these is simply the energy-momentum tensor of the
conformal theory. 
These three currents satisfy $D_- T = D_- F = D_- W=0$, 
as can be checked directly 
using the equations of motion. The expressions are consistent with the ones 
given for $D(2,1)$ in \cite{QI2} (though the basis 
used for the fields in that reference is different from the one used here). 
Note that if desired we could add a multiple of $T$ to $F$, and a
combination of derivatives of these two to $W$ to make the expressions
for $F$ and $W$ more symmetrical in the three fields, but the
resulting expressions are then slightly lengthier.

We now seek conserved currents in the affine Toda theory built on 
$D(2,1;\A)^{(1)}$
which are of the form $(J,\bar{J})$ with $J$ 
some combination of holomorphic currents from the 
conformal theory.

The energy-momentum tensor must be such a conserved current, and we can 
easily verify,  
using the equations of motion, that also $F$ leads to a conserved
current in the 
affine theory.
Explicitly, we find that (\ref{scons}) holds with
\bea
J = T ; && \bar J = 2 i \mu D_+ U_0 (\Phi) \nn
J = F ; && 
\bar J = - 2 i \mu ( \alpha D_+ \Phi_1 + (1+\alpha) D_+ \Phi_3 ) U_0
(\Phi) 
\ena
Notice that the super-energy-momentum tensor has explicitly the form
(\ref{spart}) necessary for the existence of a superpartner,
in agreement with our discussion in the last section.
For the second current based on $F$, however, the 
expression for $\bar J$ above cannot be written in the form
$D_+ K$ and we deduce that has no superpartner.

The conservation of 
$W$ does not extend to the affine theory. We find 
\be
D_- W \propto  
\left ( -D_+^3\Phi_2 + {1\over 1+\A} D_+\Phi_1 D_+^2(\Phi_2+\Phi_3)
- {1\over \A} D_+\Phi_3 D_+^2 (\Phi_1+\Phi_2) 
+ D_+\Phi_2 D_+^2 \Phi_2 \right ) U_0 (\Phi)
\ee
which cannot be written as a total $D_+$ derivative.

In order to find higher-spin conserved currents in the classical 
we look for the most general differential polynomials
in the generators $T,\,F$ and $W$ which are holomorphic in the
conformal theory. 
Note that total derivatives are trivially conserved, however, 
so we are interested only in such polynomials modulo total derivatives.
The first non-trivial candidate for a conserved current is one with 
$J = TF$, of spin 3, but this turns out not be conserved in the affine
model. 
Turning next to currents where $J$ has spin 7/2, there are 5 different
terms we can construct using $F,\,T$ and 
$W$ as building blocks (discarding total derivatives):
$$
TD_+T;\quad FD_+F;\quad TD_+F;\quad TW;\quad\mbox{and }FW
$$ 
Using these terms, we find that we can construct two conserved currents of 
spin ${7\over 2}$: 
\bea
\label{cons}
J_1 = TD_+T - {1 \over \A(1+\A)} FD_+F + 2i\,TW \nn
J_2 = TD_+F - {1+2\A \over 2\A(1+\A)} FD_+F + i\,FW
\ena
The first current reduces to the one found in \cite{QI2} when we set
$\alpha =1$, while the second has not been found previously
to our knowledge.
In the calculations we have at no point used the reality 
conditions of the fields, so the conserved currents found here are clearly 
conserved also in the twisted theory. 

To conclude this discussion of classical currents, let us summarize
the component content of what we have found.
As shown above, the spins of the superfield conserved 
currents in the $D(2,1;\A)$ 
conformal Toda model are 
$({3\over 2},{3\over 2},2)$. If we expand on component fields, 
we find 6 conserved currents 
with spins $({3\over 2},{3\over 2},2,2,2,{5\over 2})$. 
The three bosonic currents with spin 2
are modifications (by fermions) of the energy-momentum tensors of 
the three independent 
Liouville models which make up the bosonic sector of the theory. 
In the affine model, the conserved currents have spins 
$({3\over 2},{3\over 2},{7\over 2},{7\over 2})$ 
but only one of these, the full energy-momentum tensor of the theory,
is expected to have a superpartner.
If we expand on components we therefore have 
conserved currents with spins 
$({3\over 2},2,2,4,4)$. 

At first sight this seems
incompatible with the fact that the bosonic sector of the theory is composed 
of three sinh-Gordon models (for standard reality conditions), 
each of which has its own conserved energy-momentum tensor of spin 2
when the fermions are set to zero. 
However, a closer investigation of the current $W$ shows that it
contains a spin-2 part 
which {\em is}
conserved if we put the fermions to zero, even though the current $W$
itself does not satisfy a conservation equation in the full affine theory. 
Taking the expression for $D_- W$ above,
expanding in components and setting the fermions to zero we find
$$
\del_- W_2 \propto \left ( -\del_+\sigma_2 + 
{\sigma_1\over 1+\A} (\del_+\phi_2+\del_+\phi_3) - 
{\sigma_3\over\A} (\del_+\phi_1+\del_+\phi_2) + \sigma_2\del_+\phi_2 
\right ) e^{-\phi_1-\phi_2-\phi_3}
$$
where $W_2$ is the spin-2 component of $W$. Upon using the equations of motion 
for the auxiliary fields $\sigma
= \sum_j \mu \A_j e^{\A_j\cdot\phi} $ we get 
$$
\del_- W_2 \propto \del_+ \phi_2 e^{-2\phi_2}
$$
which is indeed a total derivative. 
It is interesting to note that a conserved current in a bosonic theory may 
become non-conserved in the supersymmetrized theory. 
This is reminiscent of recent results in $(1,0)$ supersymmetric 
generalizations of Toda models \cite{EM2}, 
where the addition of fermions to a bosonic conformal Toda 
theory spoils the conservation of higher-spin currents. 

\section{Quantum Conserved Currents} 
\label{qcons}

Until now, we have been considering only the classical theory.
Although we will not embark on a full-scale investigation of the
quantum theory in this paper, it is of 
considerable interest to decide whether the theory is not only 
classically integrable, but also quantum integrable. 
General arguments show that a theory with just one conserved current of 
spin higher than 2 (or in the supersymmetric case of spin higher than 3/2) is 
quantum integrable \cite{Pa}, 
and to show that the theory is quantum integrable it is 
therefore sufficient to show that the classical conserved currents found above 
are also conserved 
(with quantum corrections) in the quantum theory. 

It has been shown by Gualzetti et al \cite{QI2} that in the case of the affine 
supersymmetric Toda theory defined by $D(2,1)^{(1)}$ a quantum
conserved current
of spin 7/2 does indeed exist, and it is natural to 
expect that the corresponding current exists 
also in the deformed theory defined by $D(2,1;\A)$. However, one
cannot reject {\it a priori} the possibility that 
a quantum anomaly could spoil the conservation of this current 
for general values of $\A$,
with $\A=1$ (and $-\hf,-2$) being a  
critical value where the anomaly happens to vanishes. 
In addition, it is of interest to 
investigate the quantum behaviour of the new
current of spin-7/2 found in this paper. 
We therefore need to show explicitly 
that quantum conserved currents exist, and to do this we will apply the 
method used in \cite{QI2}. We will use standard reality conditions for the 
fields; but, as we will explain later, the choice of reality
conditions does not affect the outcome of the calculations.  

We wish to show that currents $J$ exist such that 
$D_- \langle J \rangle$ is a total $D_+$-derivative. We will use 
massless perturbation theory, i.e. we treat the interaction part of the 
Lagrangian 
\be
\label{U} 
\!\!\!U(\Phi)=\exp(-\Phi_1-\Phi_2-\Phi_3)+\exp(-\Phi_1+\Phi_2+\Phi_3)+
\exp(\Phi_1+\Phi_2-\Phi_3)+\exp(\Phi_1-\Phi_2+\Phi_3) 
\ee
as a perturbation to the free theory; for simplicity we have set $\mu$ 
equal to 
one. A potential anomaly would appear as a 
{\em local} term in 
\be
\label{expec}
D_-\langle J(X^+,X^-) \rangle = 
D_- \left \langle J(X^+,X^-) 
\exp\left ( {i\over\B^2} \int d^2 y d^2 \eta U(\Phi) \right ) 
\right \rangle_0 
\ee
where we have introduced a coupling constant $\B$ by rescaling the 
lagrangian by 
${1\over\B^2}$. 
$\langle \cdot \rangle_0$ denotes the expectation value in the free theory, 
and normal ordering of the exponential terms in $U(\Phi)$ is implicit. 
$X^\pm = (x^\pm,\theta^\pm)$ and $Y^\pm=(y^\pm,\eta^\pm)$ 
are superspace coordinates. 

Using the symmetries (\ref{sym}) of the lagrangian, any one of the four 
terms in (\ref{U}) can be transformed into any of the other terms, and 
therefore 
we need only consider one of the four exponential terms in $U(\Phi)$; 
furthermore, since we are looking for 
{\em local} terms we need expand the exponential in (\ref{expec}) only 
to first 
order. For this reason we consider
$$
D_- \left \langle J(X^+,X^-) 
\left ( {i\over\B^2} \int d^2 y d^2 \eta \exp (-\Phi_1-\Phi_2-\Phi_3)  
\right ) 
\right \rangle_0 
$$
The massless superspace propagators are 
$$
\langle \Phi_i(X) \Phi_j(Y) \rangle = 
-i\D_{ij}b_i D_-D_+ \DE(x,y) \D^{(2)}(\T-\E)
$$
where $\DE(x,y)$ is the usual bosonic propagator satisfying 
$\del_+\del_-\DE(x,y) = - {i\B^2\over 2} \D^{(2)}(x-y)$ and 
$b_i=2\e_i^2$ (the basis vectors $\e_i$ are defined in section \ref{d21}). 
The bosonic propagator is explicitly given by the expression
$$
\DE(x,y) = -{\B^2\over 4\pi} \log(2(x^+-y^+)(x^--y^-))
$$ 
with $\del_- ({1\over x^+-y^+}) = 2\pi i\D^{(2)}(x-y)$. 
It is convenient to carry out first the integration 
over the anti-commuting variables. In order to do this, we simplify the 
expressions such that each term has exactly one delta function in the 
anticommuting variables, using 
$$
D_-D_+ \D^{(2)}(\theta-\eta)=-i \quad\mbox{ and }\quad
D_-\D^{(2)}(\T-\E)D_+\D^{(2)}(\T-\E)=i\D^{(2)}(\T-\E).
$$ 
Terms which cannot be reduced to a single fermionic delta function 
using these rules will disappear. 
We do not intend to find explicit expressions for $\bar{J}$; we intend
only to show that the result is a total derivative, 
and we can therefore freely discard total derivatives to simplify 
the calculations. 

We write the current $J$ as an (implicitly normal ordered) polynomial of 
derivatives of the fields. 
It is useful to note that when considering a term which is of order 
$m+1$ in the fields, the $m$-loop calculation (i.e. $m+1$ contractions with 
the exponential)
gives a total derivative as result, and therefore 
need not be considered. Furthermore, 
a brief calculation shows that any contraction involving two fields both with 
an odd order of derivatives gives a vanishing result, since it involves higher 
powers of the delta function in the anti-commuting fields which cannot be 
removed 
using the rules given above. 

A simple calculation shows that the two classically conserved currents $T$ and 
$F$ are both conserved in the quantum theory without any quantum corrections. 
It is evident that any quantum conserved 
current is conserved also in the classical limit, so the first candidates for 
higher-spin quantum conserved currents are the 
classical spin 7/2-currents $J_i$ given in equation (\ref{cons}). 
It turns out to be most convenient to consider a completely general current of
spin 7/2, and to find coefficients such that this current is conserved in the 
quantum theory. It is then a non-trivial check of the calculations that the 
quantum currents which are found in this manner 
do indeed reduce to the known classical currents in the limit 
$\beta\rightarrow 0$. 

The most general (modulo total derivatives) current of spin 7/2, which 
respects 
the symmetries of the lagrangian, can be written as: 
\bea
J &=& k_1 D_+^3\Phi_1 D_+^4\Phi_1 + k_2D_+^3\Phi_2D_+^4\Phi_2 + 
k_3D_+^3\Phi_3D_+^4\Phi_3 +\nn
&& k_4 D_+\Phi_1D_+^3\Phi_2D_+^3\Phi_3 + k_5 D_+\Phi_2D_+^3\Phi_1D_+^3\Phi_3 +
k_6 D_+\Phi_3D_+^3\Phi_1D_+^3\Phi_2 + \nn
&& k_7 D_+^2\Phi_1D_+^2\Phi_2D_+^3\Phi_3 + 
k_8 D_+^2\Phi_2D_+^2\Phi_3D_+^3\Phi_1 +
\nn
&& k_9 D_+\Phi_1D_+^3\Phi_1 D_+\Phi_3D_+^2\Phi_3 + 
k_{10} D_+\Phi_1D_+^3\Phi_1 D_+\Phi_2D_+^2\Phi_2 +   
k_{11} D_+\Phi_2D_+^3\Phi_2 D_+\Phi_3D_+^2\Phi_3 +\nn
&&k_{12} D_+\Phi_2D_+^3\Phi_2 D_+\Phi_1D_+^2\Phi_1 +\nn
&&
k_{13} D_+\Phi_3D_+^3\Phi_3 D_+\Phi_2D_+^2\Phi_2 +
k_{14} D_+\Phi_3D_+^3\Phi_3 D_+\Phi_1D_+^2\Phi_1 +\nn
&&
k_{15} D_+\Phi_1D_+^2\Phi_1D_+^2\Phi_1D_+^2\Phi_1 + 
k_{16} D_+\Phi_2D_+^2\Phi_2D_+^2\Phi_2D_+^2\Phi_2 + 
k_{17} D_+\Phi_3D_+^2\Phi_3D_+^2\Phi_3D_+^2\Phi_3 +\nn
&& 
k_{18} D_+\Phi_1D_+^2\Phi_1D_+^2\Phi_2D_+^2\Phi_2 + 
k_{19} D_+\Phi_1D_+^2\Phi_1D_+^2\Phi_3D_+^2\Phi_3 + 
k_{20} D_+\Phi_3D_+^2\Phi_3D_+^2\Phi_2D_+^2\Phi_2 +\nn
&&k_{21} D_+\Phi_1D_+\Phi_2D_+\Phi_3D_+^2\Phi_1D_+^2\Phi_1+
k_{22} D_+\Phi_1D_+\Phi_2D_+\Phi_3D_+^2\Phi_2D_+^2\Phi_2+\nn
&&k_{23} D_+\Phi_1D_+\Phi_2D_+\Phi_3D_+^2\Phi_3D_+^2\Phi_3 
\ena
Using the observations given above, we see that we need do the 
calculations only 
up to two loops (triple contractions). 
As an example of the structure of the results of the calculations, we give 
explicitly a partial result:
\bea 
&& \hspace{-10mm}
D_- \left \langle D_+\Phi_i D_+^2 \Phi_i D_+^2 \Phi_j D_+^2\Phi_j
{i\over\B^2}\int d^2yd^2\E \exp(-\Phi_1-\Phi_2-\Phi_3) \right \rangle_0 \nn
& = & \left \{ \hf \left ( 2b_j D_+(D_+\Phi_i D_+^2\Phi_i D_+^2\Phi_j) + 
b_i D_+(D_+\Phi_i D_+^2\Phi_j D_+^2\Phi_j) + 
b_i D_+^2\Phi_i D_+^2\Phi_j D_+^2\Phi_j \right ) \right . \nn
&& -{\beta^2\over 8\pi}\left ( 
D_+^3(b_j^2D_+\Phi_i D_+^2\Phi_i + 2b_ib_j D_+\Phi_i D_+^2\Phi_j)+
D_+^2(2b_ib_j D_+^2\Phi_iD_+^2\Phi_j+b_i^2 D_+^2\Phi_j D_+^2\Phi_j) \right ) 
\nn
&& \left . +{\B^4\over 64\pi^2} 2b_ib_j \left (b_j D_+^6 \Phi_i+ 
b_i D_+^6 \Phi_j
\right ) \right \} \exp(-\Phi_1-\Phi_2-\Phi_3) + \mbox{ total derivatives} 
\ena 

After calculating the relevant part of 
$D_-\langle J \rangle$, we look for a differential polynomial 
in the fields $P(\Phi)$ of spin 5/2 such that the result of the calculation 
is equal to $D_+(P(\Phi)\exp(-\Phi_1-\Phi_2-\Phi_3))$. 
This gives a set of equations for the unknown coefficients 
$k_1$ to $k_{23}$, 
as well as for the coefficients of the polynomial $P$. 
The equations are rather cumbersome, and we have used the 
algebraic manipulation program Mathematica to solve them. 

After solving the equations, we find that two of the parameters 
$k_i$ are still 
free. One of these parameters 
gives the overall scale, but the fact that a free parameter remains 
after fixing 
the scale shows that there are 
two conserved quantum currents of spin 7/2 in the theory, 
corresponding to the two classically conserved currents 
(\ref{cons}). 
We fix the remaining two independent parameters by demanding that the currents
reduce to $J_1$ and $J_2$ in the classical limit 
(modulo total derivatives). The 
resulting expressions for the quantum conserved 
currents are rather long and are 
given in the appendix. 
In the case $\A=1$, $J_1$ corresponds to the conserved current found in 
\cite{QI2}. The classical and quantum currents denoted by $J_2$ have 
not been found before to our knowledge,
even in the case $\A=1$. 

We have not used explicitly the reality conditions of the fields in these 
calculations, and 
so the calculations are equally valid in the twisted version: if we 
define a field $\widehat{\Phi}_i$ by $\Phi_i = i \widehat{\Phi}_i$ 
and take as propagator
$$
\langle \widehat{\Phi}_i(X) \widehat{\Phi}_i(Y) \rangle = 
i b_i D_-D_+ \DE(x,y) \D^{(2)}(\T-\E)
$$
i.e. the opposite sign of the propagator of $\Phi_i$, the calculations are 
clearly not affected. 

\section{Conclusion and Discussion}
\label{con}

We have considered in this paper certain supersymmetrized 
combinations of sine/sinh-Gordon theories corresponding to 
Toda models based on the family of 
exceptional affine Lie superalgebras $D(2,1;\A)^{(1)}$ with certain
reality conditions for the fields. 
Theories based on Lie superalgebras have indefinite kinetic energy in 
general, but the models we have constructed have 
positive kinetic energy and potentials which are bounded from below. 
We have confirmed that these models are both classically and
quantum-mechanically integrable by finding explicit higher-spin
conserved currents of (super)spin 7/2.

Undoubtedly one of the most interesting features is the appearance of
the continuous parameter $\A$.
The models have a bosonic sector consisting of two sine-Gordon theories and 
one sinh-Gordon theory 
but with masses and couplings to fermions 
which depend on this continuous  
parameter.
We have also seen how all other Toda models with positive-definite
lagrangians, e.g.~the $N=2$ sine-Gordon theory, 
can be obtained as some kind of limit of this continuous
family.

Having established quantum integrability, the most 
obvious question for future study is the determination of an exact
S-matrix.
An essential step will be 
a more detailed investigation of the semi-classical 
spectrum, incorporating both sine-Gordon solitons and sinh-Gordon
fundamental particles together with the effects of supersymmetry.
We hope to return to these fascinating questions in the future. \\[5pt]

\noindent
{\bf Acknowledgements}

\noindent
The research of JME is supported by a PPARC Advanced Fellowship, while that of 
JOM is supported by the TMR European network contract number FMRX-CT96-0012, 
and is grateful to the Danish Research Council for additional support. 

\newpage
\appendix

\section{Explicit Expressions for the Quantum Conserved Currents}

The quantum conserved current which reduces to $J_1$ in the classical 
limit $\beta \rightarrow 0$ (modulo total derivatives) is:
\bea
\oo  -{1\over (1+\A)^3}
\left[ 1+{(1+\A)(1+4\A+\A^2)\over \A}\bp -35(1+\A)^2\bpp\right ] 
D_+^3\Phi_1 D_+^4\Phi_1 + \nn
\oo \left [ 1+{1-2\A-2\A^2\over \A(1+\A)}\bp-35\bpp\right ]  
D_+^3\Phi_2D_+^4\Phi_2 + \nn
\oo {1\over\A^3}\left [ 1 + {\A(-2-2\A+\A^2)\over 1+\A}\bp -35\A^2\bpp 
\right ] 
D_+^3\Phi_3D_+^4\Phi_3 +\nn
\oo {2(1-\A)\over \A^2}\left [ 1-{3\A\over 1+\A}\bp\right ] 
D_+\Phi_1D_+^3\Phi_2D_+^3\Phi_3 
 -{2(1+2\A)\over \A^2(1+\A)^2}\left [ 1 - 3\A(1+\A)\bp \right ] 
D_+\Phi_2D_+^3\Phi_1D_+^3\Phi_3 +\nn
\oo -{2(2+\A)\over (1+\A)^2} \left [ 1-{3(1+\A)\over \A}\bp\right ] 
D_+\Phi_3D_+^3\Phi_1D_+^3\Phi_2 + \nn
\oo {4(1-\A)\over \A(1+\A)}\bp D_+^2\Phi_1D_+^2\Phi_2D_+^3\Phi_3 + 
{4(2+\A)\over \A(1+\A)} \bp D_+^2\Phi_2D_+^2\Phi_3D_+^3\Phi_1 + \nn
\oo {1-\A\over \A(1+\A)^2}
\left [ 1 + {12(1+\A)\over 1-\A}\bp\right ] 
D_+\Phi_1D_+^3\Phi_1 D_+\Phi_3D_+^2\Phi_3 + \nn
\oo -{1-\A\over \A(1+\A)^2} \left [ 1 - {12\A(1+\A)\over 1-\A}\bp \right ] 
D_+\Phi_1D_+^3\Phi_1 D_+\Phi_2D_+^2\Phi_2  + \nn
\oo {1+2\A\over \A(1+\A)}\left [1-{18(1+\A)\over 1+2\A}\bp\right ] 
D_+\Phi_2D_+^3\Phi_2 D_+\Phi_3D_+^2\Phi_3   +\nn
\oo -{1+2\A\over \A(1+\A)}\left [1-{18\A\over 1+2\A}\bp\right ]
D_+\Phi_2D_+^3\Phi_2 D_+\Phi_1D_+^2\Phi_1 +\nn
\oo {2+\A\over \A^2(1+\A)} \left [ 1-{12\A(1+\A)\over 2+\A}\bp \right ] 
D_+\Phi_3D_+^3\Phi_3 D_+\Phi_2D_+^2\Phi_2 +\nn
\oo -{2+\A\over \A^2(1+\A)} \left [ 1-{18\A\over 2+\A}\bp \right ] 
D_+\Phi_3D_+^3\Phi_3 D_+\Phi_1D_+^2\Phi_1 +\nn
\oo-{1\over (1+\A)^3} \left [ 1-3(1+\A)\bp\right ]
D_+\Phi_1D_+^2\Phi_1D_+^2\Phi_1D_+^2\Phi_1 + \nn
\oo \left [ 1+3\bp\right ] D_+\Phi_2D_+^2\Phi_2D_+^2\Phi_2D_+^2\Phi_2 + 
{1\over\A^3} \left [ 1+3\A\bp \right ] 
D_+\Phi_3D_+^2\Phi_3D_+^2\Phi_3D_+^2\Phi_3 +\nn
\oo - {6\over 1+\A}\bp D_+\Phi_1D_+^2\Phi_1D_+^2\Phi_2D_+^2\Phi_2 - 
{6\over\A(1+\A)}\bp D_+\Phi_1D_+^2\Phi_1D_+^2\Phi_3D_+^2\Phi_3 + \nn
\oo {6\over \A}\bp D_+\Phi_3D_+^2\Phi_3D_+^2\Phi_2D_+^2\Phi_2 +\\
\oo \frac{2}{\A(1+\A)} D_+\Phi_1D_+\Phi_2D_+\Phi_3
\left [ {1-\A\over 1+\A} D_+^2\Phi_1D_+^2\Phi_1-
(1+2\A)D_+^2\Phi_2D_+^2\Phi_2+ {2+\A\over\A} D_+^2\Phi_3D_+^2\Phi_3 \right ], 
\nonumber 
\ena
\newpage 
while the current which reduces to $J_2$ in the classical limit is:
\bea
\oo -\left [{\A\over (1+\A)^3}+{(1+5\A)\over \A(1+\A)^2}\bp 
+{5(3-10\A)\over(1+\A)} \bpp\right ] 
D_+^3\Phi_1 D_+^4\Phi_1 + \nn 
\oo -\left [{(1+2\A)(-1+3\A+3\A^2)\over \A(1+\A)}\bp-15(1+2\A)\bpp\right ]  
D_+^3\Phi_2D_+^4\Phi_2 + \nn
\oo \left [ {1+\A\over\A^3} + {4+5\A \over \A^2(1+\A)}\bp -
{5(13+10\A)\over\A}\bpp \right ] 
D_+^3\Phi_3D_+^4\Phi_3 +\nn
\oo \left [ {2\over\A^2}+{6(1+\A+2\A^2)\over \A(1+\A)}\bp - 
192 \bpp \right ] 
D_+\Phi_1D_+^3\Phi_2D_+^3\Phi_3 + \nn
\oo -\left [ {2(1+3\A(1+\A))\over \A^2(1+\A)^2}+ {6\over\A(1+\A)}\bp -
192\bpp \right ] 
D_+\Phi_2D_+^3\Phi_1D_+^3\Phi_3 + \nn
\oo \left [ {2\over (1+\A)^2} +{6(2+3\A+2\A^2)\over \A(1+\A)}\bp - 
192\bpp \right ] 
D_+\Phi_3D_+^3\Phi_1D_+^3\Phi_2 + \nn
\oo {4(1+3\A)\over \A(1+\A)}\bp D_+^2\Phi_1D_+^2\Phi_2D_+^3\Phi_3 + 
{4(2+3\A)\over \A(1+\A)} \bp D_+^2\Phi_2D_+^2\Phi_3D_+^3\Phi_1 + \nn
\oo \left [ {1+3\A\over \A(1+\A)^2} + {12\over\A}\bp\right ] 
D_+\Phi_1D_+^3\Phi_1 D_+\Phi_3D_+^2\Phi_3 
- {1+3\A\over \A(1+\A)^2} D_+\Phi_1D_+^3\Phi_1 D_+\Phi_2D_+^2\Phi_2  + \nn
\oo\left [ {1\over \A(1+\A)}-{6(3+2\A)\over\A}\bp\right ] 
D_+\Phi_2D_+^3\Phi_2 D_+\Phi_3D_+^2\Phi_3 + \nn  
\oo-\left [ {1\over \A(1+\A)}+{6(1-2\A)\over 1+\A}\bp\right ]
D_+\Phi_2D_+^3\Phi_2 D_+\Phi_1D_+^2\Phi_1 +\nn
\oo
{2+3\A\over \A^2(1+\A)} D_+\Phi_3D_+^3\Phi_3 D_+\Phi_2D_+^2\Phi_2 
 -\left [ {2+3\A\over \A^2(1+\A)} -{6(1+4\A)\over\A(1+\A)}\bp \right ] 
D_+\Phi_3D_+^3\Phi_3 D_+\Phi_1D_+^2\Phi_1 +\nn
\oo
-\left[ {\A\over (1+\A)^3} + {3(1-2\A)\over (1+\A)^2} \bp\right ]
D_+\Phi_1D_+^2\Phi_1D_+^2\Phi_1D_+^2\Phi_1 + \nn
\oo - 3(1+2\A) \bp  D_+\Phi_2D_+^2\Phi_2D_+^2\Phi_2D_+^2\Phi_2 + 
 \left [ {1+\A \over\A^3} +{3(3+2\A)\over\A^2} \bp \right ] 
D_+\Phi_3D_+^2\Phi_3D_+^2\Phi_3D_+^2\Phi_3 +\nn
\oo 
+ {6\over 1+\A}\bp D_+\Phi_1D_+^2\Phi_1D_+^2\Phi_2D_+^2\Phi_2 - 
{6(1+2\A)\over\A(1+\A)}\bp D_+\Phi_1D_+^2\Phi_1D_+^2\Phi_3D_+^2\Phi_3 + \nn
\oo {6\over \A}\bp D_+\Phi_3D_+^2\Phi_3D_+^2\Phi_2D_+^2\Phi_2 +
\left [ {2(1+3\A)\over \A(1+\A)^2} - {24\over 1+\A}\bp \right ] 
D_+\Phi_1D_+\Phi_2D_+\Phi_3 D_+^2\Phi_1D_+^2\Phi_1 + \nn
\oo - \left [ {2\over\A(1+\A)} - 24 \bp \right ] 
D_+\Phi_1D_+\Phi_2D_+\Phi_3 D_+^2\Phi_2D_+^2\Phi_2 + \nn
\oo  \left [ {2(2+3\A)\over\A^2(1+\A)} + {24\over\A} \bp \right ] 
D_+\Phi_1D_+\Phi_2D_+\Phi_3 D_+^2\Phi_3D_+^2\Phi_3
\ena


\newpage
\begin{thebibliography}{99}

\bibitem{EM} J.M. Evans and J.O. Madsen, \NPB{503} (1997) 715. 

\bibitem{To} F. Toppan, {\it Susy hierarchies and affine algebras}, 
solv-int/9710001. 

\bibitem{Dict}
V.G. Kac, Adv. Math. {\bf 26} (1977) 8; \newline
L. Frappat, A. Sciarrino and P. Sorba,  
{\it Dictionary on Lie Superalgebras}, \newline 
hep-th/9607161; preprint ENSLAPP-AL-600/96, DSF-T-30/96 

\bibitem{JE}{J.M. Evans, \NPB{390} (1993) 225 \newline
J.M. Evans and J.O. Madsen, \PLB{384} (1996) 131}

\bibitem{Ra}
R.F. Dashen, B. Hasslacher and A. Neveu, Phys. Rev. {\bf D11} (1975) 3424; 
\newline 
R. Rajaraman: {\it Solitons and Instantons}, Elsevier Science B.V. 1982. 

\bibitem{FSS}L. Frappat, A. Sciarrino and P. Sorba,\CMP{121} (1989)
457. 

\bibitem{QI2}
A. Gualzetti, S. Penati and D. Zanon, \NPB{398} (1993) 622

\bibitem{EH} J. Evans and T. Hollowood, \PLB{293} (1992) 100. 

\bibitem{EHM}{J.M. Evans, N.J. MacKay and M. Hassan,
{\it Conserved charges and supersymmetry in principal chiral models},
hep-th/9711140.}

\bibitem{SHR}
L. Frappat, E. Ragoucy and P. Sorba, \CMP{157} (1993) 499

\bibitem{EM2}{J.M. Evans and J.O. Madsen, \PLB{389} (1996) 665}

\bibitem{Pa} S. Parke, \NPB{174} (1980) 166. 

\end{thebibliography}
\end{document}